\documentclass[useAMS,usenatbib,usegraphicx]{mn2e}
\usepackage{graphicx}                                            
\usepackage{times}
\usepackage{float}
\usepackage{rotating}
\usepackage{epstopdf}
\usepackage{multirow}

\newcommand{\ergcms}{erg~cm$^{-2}$~s$^{-1}$}
\newcommand{\ergps}{erg~s$^{-1}$}

\def\ltsima{$\; \buildrel < \over \sim \;$}
\def\simlt{\lower.5ex\hbox{\ltsima}}
\def\gtsima{$\; \buildrel > \over \sim \;$}
\def\simgt{\lower.5ex\hbox{\gtsima}}
\def\gsimeq
{\hbox{\raise0.5ex\hbox{$>\lower1.06ex\hbox{$\kern-1.07em{\sim}$}$}}}
\def\lsimeq
{\hbox{\raise0.5ex\hbox{$<\lower1.06ex\hbox{$\kern-1.07em{\sim}$}$}}}

\def\xmm{{\it XMM-Newton }}

\def\xmm{{\it XMM-Newton}}
\def\chandra{{\it Chandra}}
\def\suzaku{{\it Suzaku}}

\def\nustar{{\it NuSTAR}}
\def\swift{{\it Swift}}
\def\integral{{\it INTEGRAL}}

\def\apj{ApJ}
\def\mnras{MNRAS}
\def\aap{A\&A}
\def\apjl{ApJ}
\def\apjs{ApJS}
\def\araa{ARA\&A}
\def\pasj{PASJ}
\def\nat{Nature}
\def\iaucirc{IAU~circular}
\def\procspie{Proc. SPIE}
\def\gca{GeCoA}
\def\ssr{SSRv}

\def\exo{EXO~0748-676}
\def\axj{AX~J1745.6-2901}
\def\sgras{Sgr~A$^{\star}$}

\def\Fevc{Fe~{\sc xxv}}
\def\Fevs{Fe~{\sc xxvi}}

\def\xis{XIS}
\def\xis1{XIS1}
\def\xis2{XIS2}
\def\xis3{XIS3}

\title[] 
 {{On the Fe~K absorption - accretion state connection in the Galactic center
 neutron star X-ray binary AX~J1745.6-2901}}

 \author[G.\ Ponti et al. ]
 {G.~Ponti$^{1}$, S.~Bianchi$^{2}$, T.~Mu\~{n}oz-Darias$^{3}$, B.~DeMarco$^{1}$, 
 T.~Dwelly$^{1}$, R.~P.~Fender$^{3}$, 
 \newauthor
  K. Nandra$^{1}$, N.~Rea$^{4,5}$, K.~Mori$^{6}$, D.~Haggard$^{7}$, 
  C.~O.~Heinke$^{8}$, N.~Degenaar$^{9}$, T. Aramaki$^{6}$ 
 \newauthor
  M.~Clavel$^{10,11}$,
  A.~Goldwurm$^{10,11}$, C.~J.~Hailey$^{6}$,  G.~L.~Israel$^{12}$, 
  M.~R.~Morris$^{13}$, 
 \newauthor
 A.~Rushton$^{3,14}$ and R.~Terrier$^{10}$ 
\\
   $^1$ Max-Planck-Institut f{\"u}r Extraterrestrische Physik, Giessenbachstrasse, D-85748, Garching, Germany\\
   $^2$ Dipartimento di Matematica e Fisica, Universit\`a Roma Tre, Via della Vasca Navale 84, I-00146, Roma, Italy\\
   $^3$ University of Oxford, Department of Physics, Astrophysics, Denys Wilkinson Building, Keble Road, Oxford, OX1 3RH, UK\\
   $^4$ Astronomical Institute Anton Pannekoek, University of Amsterdam, Postbus 94249, NL-1090 GE Amsterdam, the Netherlands\\
   $^5$ Institute of Space Sciences (CSIC-IEEC), Campus UAB, Torre C5, 2a planta, E-08193 Barcelona, Spain\\
   $^6$ Columbia Astrophysics Laboratory, Columbia University, New York, NY 10027, USA\\
   $^7$ CIERA and Physics and Astronomy Department, Northwestern University, 2145 Sheridan Road, Evanston, IL 60208, USA\\
   $^8$ Department of Physics, University of Alberta, CCIS 4-183, Edmonton, AB T6G 2E1, Canada\\
   $^9$ Department of Astronomy, University of Michigan, 500 Church Street, Ann Arbor, MI 48109, USA\\
   $^{10}$ AstroParticule et Cosmologie, Universit\'e Paris Diderot, CNRS/IN2P3, CEA/DSM, Observatoire de Paris, Sorbonne Paris Cit\'e, 10 rue Alice Domon \\ et L\'eonie Duquet, 75205 Paris Cedex 13, France\\
   $^{11}$ Service d' Astrophysique/IRFU/DSM, CEA Saclay, B\^at. 709, 91191 Gif-sur-Yvette Cedex, France\\
   $^{12}$ INAF - Osservatorio astronomico di Roma, Via Frascati 44, Monteporzio I-00040, Catone (Roma), Italy\\
   $^{13}$ Department of Physics \& Astronomy, University of Calisfornia, Los Angeles, CA 90095-1547, USA\\
   $^{14}$ School of Physics and Astronomy, University of Southampton, Highfield, Southampton SO17 1BJ, UK
}
\pagerange{\pageref{firstpage}--\pageref{lastpage}}
\usepackage{times}
\begin{document}
\label{firstpage}
 \maketitle
\begin{abstract}
\axj\ is a high-inclination (eclipsing) neutron star Low Mass X-ray
Binary (LMXB) located less than $\sim1.5$~arcmin from Sgr~A$^\star$.
Ongoing monitoring campaigns have targeted Sgr~A$^\star$ frequently 
and these observations also cover \axj.  We present here an X-ray analysis of
\axj\ using a large dataset of 38 \xmm\ observations, including
eleven which caught \axj\ in outburst.  Fe~K absorption is clearly
seen when \axj\ is in the soft state, but disappears during the hard
state. The variability of these absorption features does not
appear to be due to changes in the ionizing continuum.
The small K$\alpha$/K$\beta$ ratio of the equivalent widths of the \Fevc\ 
and \Fevs\ lines suggests that the column densities and turbulent velocities of the
absorbing ionised plasma are in excess of $N_H\simeq 10^{23}$~cm$^{-2}$
and $v_{\rm turb}\gsimeq500$~km~s$^{-1}$.
These findings strongly support a connection between the wind (Fe~K absorber) 
and the accretion state of the binary. These results reveal strong similarities between 
\axj\ and the eclipsing neutron star LMXB, EXO~0748-676, as well as 
with high-inclination black hole binaries, where winds (traced by the same Fe~K 
absorption features) are observed only during the accretion-disc-dominated 
soft states, and disappear during the hard states characterised by jet emission.
\end{abstract}


\section{Introduction}

Equatorial accretion disc winds have recently been demonstrated to be an 
ubiquitous feature of accreting black holes (Ponti et al. 2012). Such winds 
have almost always been observed during the soft state and typically disappear 
during the canonical hard state (Neilsen \& Lee 2009; Ponti et al. 2012; 
Miller et al. 2012). 
The estimated wind mass outflow rates (Lee et al. 2002; Ueda et al. 2004; 
Miller et al. 2006; Neilsen et al. 2011; Ponti et al. 2012; King et al. 2012) and 
their tight connection with the accretion state suggest that winds are a fundamental 
component of the accretion process in black hole binaries. 

Accreting Neutron Stars (NS) are also known to have equatorial absorbers
and winds (Ueda et al. 2001; Sidoli et al. 2001; Parmar et al. 2002;
Boirin et al. 2003; 2004; 2005; Diaz-Trigo et al. 2006; 2013, but see also 
Miller et al. 2011) however a one-to-one connection between the wind and 
accretion state has still to be established. 
A recent study focused on the absorption properties 
of the neutron star low-mass X-ray binary \exo\ (Ponti et al. 2014). 
This source has been continuously in outburst for 23 years (it was discovered 
in 1985 and it returned to quiescence in 2008; Parmar et al. 1985; 
Wolff et al. 2008; Hynes \& Jones 2008). 
As characteristic of the high-inclination sources, \exo\ shows dips and 
eclipses. An inclination of $75<i<83^{\circ}$ was estimated (Parmar et 
al. 1985; 1986) for a primary mass of $M_{NS}\sim1.4$~M$_{\odot}$, 
which is consistent with dynamical estimates (Mu\~{n}oz-Darias et al. 
2011; Ratti et al. 2012). 
No Fe~K absorption lines are detected during the more than
20 \xmm\ observations which catch \exo\ in the hard state. Nonetheless, 
intense Fe~{\sc xxiii}, \Fevc\ and \Fevs\ absorption lines are clearly
observed in the single \xmm\ observation where \exo\ is seen in the
soft state (Ponti et al. 2014). This suggests that the wind--accretion
state connection might also be present in some accreting neutron star 
binaries. To further test this hypothesis, we analyse the
\xmm\ and \swift\ observations (as well as \nustar\ data to constrain the 
broad band Spectral Energy Distribution; SED) of another well known 
high-inclination (dipping and eclipsing) accreting neutron star, \axj.

At less than 1.5 arcminutes from Sgr~A$^\star$, \axj\ lies within 
one of the most intensely observed patches of the X-ray sky.  
\axj\ was identified as a new transient X-ray burster near 
the Galactic center in 1993--1994 and 1997 ASCA observations 
(Maeda et al. 1996; Kennea et al. 1996; Sakano et al. 2002). 
Intensity dips with a period of $8.356\pm0.008$ hours were identified, 
indicating the high-inclination of the source. Excess soft X-rays 
during the dips are attributed to scattering by interstellar dust (Maeda et al. 1996).

Muno et al. (2003) catalogued a faint 
($L_X\sim10^{32}$ \ergps, for an 8 kpc distance) Chandra X-ray source, 
CXOGC J174535.6-290133, confirmed to be the quiescent X-ray counterpart 
of \axj\ (Heinke et al. 2008). New transient outbursts were 
seen by \swift, \integral, \xmm, \chandra\ and \suzaku\ in early 2006 
(Kennea et al. 2006; Chenevez et al. 2006), in 2007--2008 
(Wijnands et al. 2007; Porquet et al. 2007; Degenaar et al. 2009; 
Hyodo et al. 2009; Degenaar et al. 2010a), in 2010 (Degenaar 2010b), 
and 2013-2014 (Degenaar et al. 2013a; 2014a). The 2006 and 2010 outbursts 
were short (4 and 4--7 months respectively\footnote{The end date of the 2010 
outburst is uncertain as it occured when \axj\ was too close to the Sun to be 
monitored by \swift\ or any other X-ray telescope.}) and low luminosity (peak 
$L_X\sim5\times10^{35}$~\ergps), while the 2007--2008 and 2013--2014 
outbursts were longer (1.5--1.7 years, and $>$1 year to date) and of higher 
luminosity (peak $L_{2-10~keV}\sim7\times10^{36}$ \ergps\ for both; 
Degenaar et al. 2010a; 2014a). 

As observed in many other high-inclination systems (e.g. Boirin et al. 2004; 
Diaz-Trigo et al. 2006; 2013), \axj\ shows eclipses, dips, and Fe K 
absorption (both \Fevc\ and \Fevs\ K$\alpha$ and K$\beta$ absorption; 
Hyodo et al. 2009). The observed equivalent widths of these absorption 
lines range from $\sim$30 to 60 eV, and the lines are observed during all 
orbital phases (except eclipses). Therefore, a disc corona origin for the 
absorbing material has been proposed (Hyodo et al. 2009). Hyodo et al. 
(2009) suggested that the absorbing gas is outflowing with a bulk motion 
of $\sim10^3$ km s$^{-1}$, and also showed that the dip spectra are 
well-reproduced by increased absorption by cold (approximately neutral) material.

A total of 38 \xmm\ observations included \axj, of which 
11 have caught the source in outburst.  The detailed Swift monitoring of the 
Galactic center (Degenaar et al. 2009; 2010; 2013), comprising over 1000 \swift\ 
snapshot observations obtained between 2006 and 2014, allows us to place 
the \xmm\ observations in context, tracking the accretion state of \axj. 
Taken together, these data provide us with a unique opportunity to study 
the wind in \axj\ and determine whether the appearance/disappearance 
of the wind is linked to the accretion state of the source. 

The paper is organised as follows. In Section 2 we present the \xmm\
and \swift\ observations and data reduction methods. In Section 3 we
present the method used to determine the accretion state.  In Sections
4 and 5 we present detailed modelling of the \xmm\ data using
phenomenological and proper photo-ionisation models.  Our results are
summarised in Section 6. 

\section{Observations and data reduction}

All spectral fits were performed using the {\sc Xspec} software package (version 12.7.0). 
Uncertainties and upper limits are reported at the 90 per cent confidence level for 
one interesting parameter, unless otherwise stated. 
The reported X-ray fluxes are not corrected for Galactic absorption. 
To allow the use of $\chi^2$ statistics we group each spectrum to have 
a minimum of 25 counts in each bin. We adopt a nominal Eddington limit for \axj\ of  
L$_{Edd}=2\times10^{38}$~\ergps (appropriate for a primary mass 
of $M_{NS}\sim1.4$~M$_{\odot}$ and cosmic composition; Lewin et al. 1993).

\subsection{\xmm}
\label{xmmDR}

Several independent groups, with a wide variety of science goals have made \xmm\ 
observations of the Galactic Centre field over the last decade.
In this paper we combine all available data where \axj\ lies within the \xmm\ 
field of view. This includes recent \xmm\ observations which were designed 
to monitor the passage of G2 (Gillessen et al. 2012) at peri-center (PIs: Haggard; 
Ponti), and to track the evolution of the outburst of SGR~J1745-2900 
(Mori et al. 2013; Kennea et al. 2013; Rea et al. 2013; PI: Israel), data from the 
\xmm\ scan of the central molecular zone (PI: Terrier), plus many older archival 
\xmm\ observations (See Tab. \ref{data} and Tab. 3 and 5 of Ponti et al. in prep.). 

As of 2014 May 14 there were 34 observations publicly 
available in the \xmm\ archive (Tab. \ref{data}), pointed near \axj\
and with EPIC-pn clean exposure longer than 3~ks. We add to this four 
new proprietary observations that were accumulated between August 2013 
and April 2014.

Starting from the \xmm\ Observation Data Files (ODF), we re-process all the data sets, 
with the latest version (13.5.0) of the \xmm\ Science Analysis System (SAS), 
applying the most recent (as of 14/05/2014) calibrations. 
Because of the relatively small effective area of the MOS cameras in the Fe~K band, 
we restrict our analysis here to data collected with the EPIC-pn camera. 

The majority of the EPIC-pn observations have been accumulated in Full
Frame mode with the medium filter. One observation ({\sc obsid} 0112972101) was 
performed in Extended Full Frame mode, and two ({\sc obsid} 0111350301 
and 0111350101) used the thick filter.

Photon pile-up affected all observations in which \axj\ was found in the soft state
(see Section 3; Tab. \ref{data}). To mitigate the effects of pile-up
on the extracted spectra in the soft state, we adopt an annular extraction region
centred on the source, with inner radius of $r_{\rm in}=9.25$~arcsec and
outer radius of $r_{\rm out}=45$~arcsec (see e.g. van Peet et al. 2009).
Discarding all the photons within $9.25$~arcsec of the peak of the source 
point-spread function (PSF), we remove $\sim50~\%$ of the encircled 
energy from the source which is concentrated on a small area (see fig. 7 
in section 3.2.1.1 of the \xmm\ users' handbook
\footnote{http://xmm.esac.esa.int/external/xmm\_user\_support/documentation/uhb/index.html}). This allows us to 
reduce the effects of pile-up significantly. An outer radius of $45$~arcsec allows us to retain 
$\sim95~\%$ of the remaining source photons, whilst keeping the instrumental background and diffuse 
emission low. 
For the hard state observations, which are not affected by pile-up, 
we extracted events using a circular region with $45$~arcsec radius. 
To compute the source flux during 
the quiescent (or close to quiescence) observations whilst minimising the contamination 
from instrumental background and diffuse 
emission, we used a small circular extraction 
region with a radius of $10$~arcsec (see Tab. \ref{data}).
 
We initially selected the background photons from
a region of similar size and shape and on the same detector chip as
the source region. However, due to the bright and highly
inhomogeneously distributed diffuse emission near the Galactic center
(Wang et al. 2002; Baganoff et al. 2003; Koyama et al. 2007; Ponti et al. 2010; 
2013; Clavel et al. 2013),
we decided to accumulate the background spectrum from the same region of the 
sky selected for the source during the \xmm\ observations where \axj\ was 
in quiescence (see Tab. \ref{data}). 
To mitigate the effects of the spatial and long term temporal dependence of 
the internal EPIC-pn particle background, we selected only the quiescent 
state observations in which \axj\ was near the optical axis of the EPIC-pn 
instrument and which were taken after September 2006 (corresponding to the 
first observation in which \axj\ was observed in outburst by \xmm, see Tab. \ref{data}). 
We identified periods of enhanced particle-induced background activity by calculating the 
full detector light curve in the 12--15~keV band, after excluding the events within a 
2.5~arcmin radius of \axj. Excluding all photons within 2.5~arcmin from the source ensures 
that more than $95~\%$ of the photons from \axj\ are removed and that the emission from 
\sgras\ and its immediate surrounding are also excluded. 
Time intervals with a count rate higher than the threshold specified in Tab. \ref{data} 
were filtered out and not considered in further analysis. 

In order to identify and remove type I bursts from the analysis we used 
a 3~s resolution hard-band X-ray light curve. The 5--10~keV band 
was chosen for this as it is only marginally affected by dipping (Diaz-Trigo et 
al. 2006; van Peet et al. 2009; Ponti et al. 2014).
We identified the intervals where \axj\ was in eclipse using a 60s binned 5--10~keV light curve. 
The thresholds we applied are reported in Tab. \ref{data}. 

Absorption dips are generally revealed by sudden increases in the
hardness-ratio. Therefore, to investigate the dipping phenomenon we examined
the hardness-ratio lightcurve, defined here as the ratio of the 
$5-10$~keV light curve to the $0.5-5$~keV light
curve. Following van Peet et al. (2009) and Ponti et al. (2014) we
determined the average hardness-ratio for those intervals clearly
belonging to the persistent emission, and then flagged as
\textit{dipping} those periods having hardness-ratio 1.5 times larger
than the persistent value. We note that the light curves of the source
are only moderately affected by the dipping phenomenon. 
However this might be due to the high
foreground absorption column density towards \axj, which prevents us
from directly studying the soft X-ray energies where dips are most
easily revealed.

After applying the particle background cut and the removal of bursts
and eclipses, we extracted, for each observation, a source spectrum of the 
persistent emission plus mild dipping (e.g. dips with hardness ratio lower 
than 1.5 time quiescence; see Tab. \ref{data} and Ponti et al. 2014). 
For each spectrum, the response matrix and effective area have been 
computed with the XMM-SAS tasks {\sc rmfgen} and {\sc arfgen}.

\begin{figure*}
\includegraphics[width=0.9\textwidth,height=0.65\textwidth]{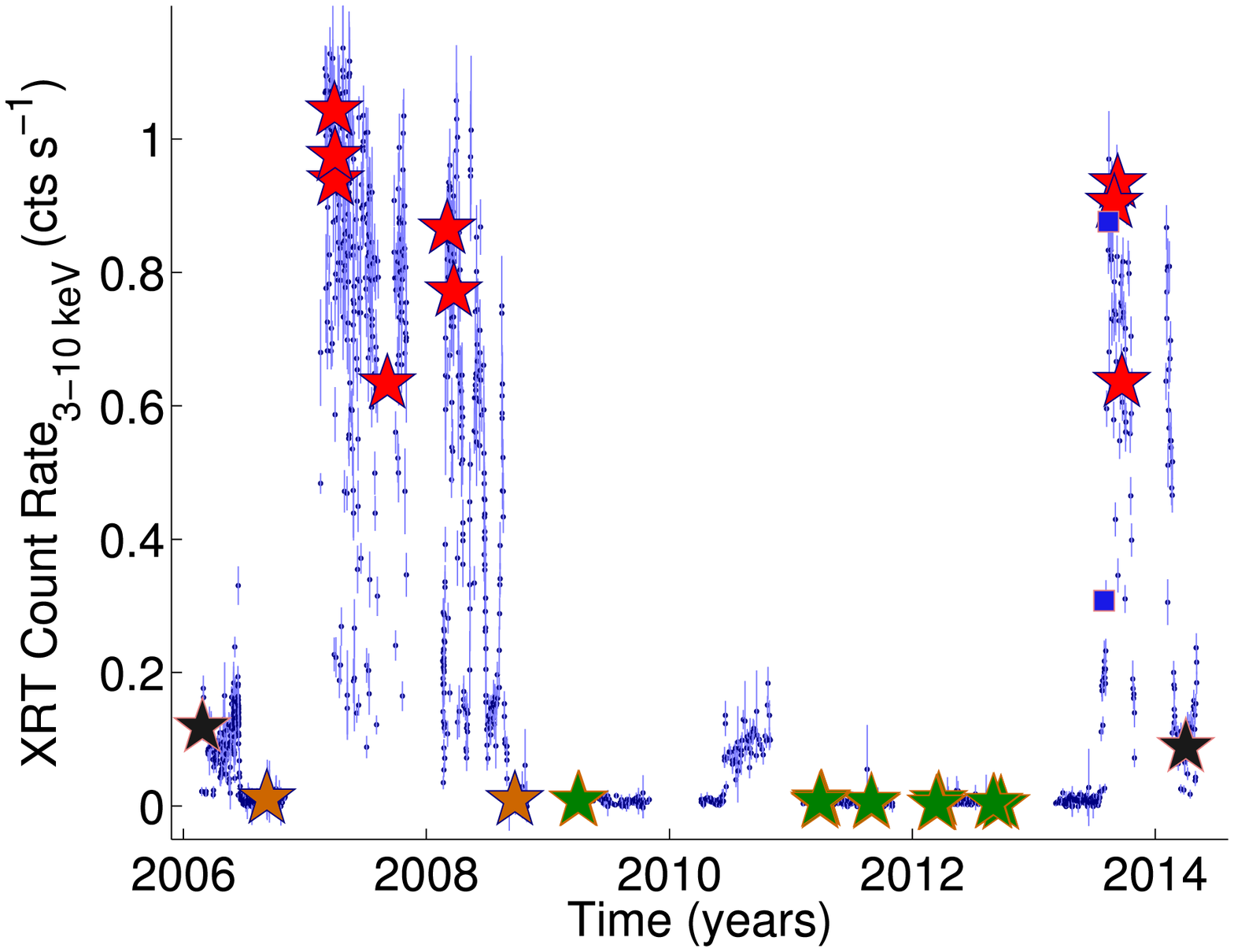}
\caption{The X-ray light curve of \axj, spanning $>8$ years of \swift, \xmm\ and \nustar\ 
observations. Each small blue point shows the normalised 3--10~keV count rate 
for each \swift\ visit. The large stars indicate the \xmm\ EPIC-pn countrate, 
(converted to the equivalent \swift-XRT count rate) measured during the observations 
which caught \axj\ in the quiescent (green), hard (black) and soft (red) states. 
The blue squares indicate the observed count rate (converted to the equivalent 
\swift-XRT count rate) during the hardest and softest \nustar\ observations, which were 
used to constrain the source's broad-band Spectral Energy Distribution (SED; see 
Section \ref{SEDmod}). 
The \xmm\ observations before 2006 all found \axj\ in the quiescent state and so are not shown.
}
\label{LC}
\end{figure*}

\subsection{\nustar}

Several \nustar\ observations of the Sgr~A$^\star$ field have been carried
out to date, with the primary science goal of monitoring the time
evolution of the newly discovered magnetar SGR~J1745-29 (Kaspi et
al. 2014). Four of these observations caught \axj\ during its 2013
outburst (which began on 2013~July~25).  These four observations
were carried out on 2013~July~31, August~8, 9 and 13, and have
exposure times 22.3, 12.0, 11.2 and 11.7~ks respectively (\nustar\
{\sc obsid}s 80002013018, 80002013020, 80002013022 and 80002013024).
No further \nustar\ observations of the Galactic Center were
performed until June 2014.  In each of the four \nustar\
observations \axj\ was the brightest source in the
field of view, yielding adequate photon statistics to study spectral
evolution over time. In this paper, \nustar\ data were used only to 
constrain the hardest and softest broad-band Spectral Energy
Distributions (SED) exhibited by \axj. Full details of the outburst evolution and an
analysis of type-I X-ray bursts will be presented by Hailey et
al. (in prep.), but here we briefly describe our \nustar\ data analysis.

All \nustar\ data processing was performed with the {\it \nustar\ Data Analysis 
Software (NuSTARDAS)} v.1.3.1. \nustar\ consists of two co-aligned X-ray 
telescopes (FPMA and FPMB) with an energy range 3--79~keV and having 
spectral resolution 400~eV (FMHM) at 10~keV (Harrison et al. 2013). 
We first filtered out time intervals containing type-I X-ray bursts in a similar 
way to that carried out for the \xmm\ data, then extracted source photons 
from a $30$~arcsec radius circle around the source position of \axj. 
This extraction region is well-calibrated for a point source, and results in 
negligible contamination from the nearby magnetar SGR~J1745-29 
($\sim1.5$~arcmin away from \axj). 
We extracted background spectra from a pre-outburst \nustar\ 
observation taken on 2013~July~7 ({\sc obsid} 8002013016, i.e. the last observation before 
the outburst) using the source extraction region. 
We re-binned the \nustar\ spectra to minimum 30 counts per bin, and discarded events 
outside the nominal \nustar\ bandpass (3--79~keV). 
We found that the softest and hardest spectra of \axj\
were observed in {\sc obsid}s 80002013018 and 80002013024 respectively.
We fit the softest observation (FPMA and FPMB spectra jointly) using the XSPEC 
model {\sc phabs*simpl*diskbb} (see Steiner et al. 2009 for more details about 
the {\sc simpl} Comptonization component). We use the same model to fit the hard state observation 
even if the peak of the disc blackbody component is not in the \nustar\ band (see \S 
\ref{SEDmod}). The fit parameters were later adopted to construct 
the input SED data for our photo-ionization models (Section 5). 

\subsection{\swift}

The \swift\ telescope has been used to regularly monitor the Galactic center
field since February 2006, typically making one or more short ($\sim1$~ks)
snapshot observations on each day when the Galactic centre is visible (Degenaar et al. 2013b).
Our reduction of the \swift-XRT (X-Ray Telescope, Burrows et al. 2000) data 
utilises a pipeline developed for monitoring campaigns of bright AGN (see Fabian et al. 2012, 
Cameron et al. 2012, Connolly et al. 2014).
We have extracted an X-ray light-curve for \axj\ using all
\swift-XRT `photon counting' mode observations with a nominal aim point
within 25~arcmin of Sgr~A$^\star$. For the work presented in this paper 
we have analysed a total of 1116 \swift\ {\sc obsid}s, including over 2000 separate 
visits with a summed exposure time of 1226~ks.

The raw \swift-XRT datasets were downloaded from the HEASARC archive and 
reprocessed using the tool {\sc xrtpipeline} (version 0.12.8).
For each \swift\ visit we used {\sc xselect} to extract source events from a
30~arcsec radius circular aperture centred on 17:45:35.65, -29:01:34.0 (J2000). 
Background events were extracted from a nearby 60~arcsec radius circular region 
containing no bright sources (centre 17:45:28.90, -29:03:44.2, J2000). 
We measured the number of detected counts in the source and background 
regions for each of a standard set of ten medium-width energy bands
spanning the range 0.3--10~keV. The background-subtracted
source count-rate in each energy band was calculated from the observed
counts taking account of the relative sky areas of the source and
background regions.

The sensitivity of \swift-XRT varies across the focal plane due to
vignetting and the presence of bad pixels. We have compensated for the
varying instrumental sensitivity between visits by calculating
visit-specific corrections using the following method. For each visit
we generated an {\sc arf} file using the standard \swift\ tools ({\sc xrtexpomap} 
and {\sc xrtmkarf}). Using this visit-specific {\sc arf}, the nominal XRT response 
matrix, a simple absorbed power-law spectral model and the {\sc fakeit} function within {\sc xspec}, we
calculated the expected count-rate for the visit-specific source
extraction aperture ($R_{\mathrm{fake},i}$) and the count-rate
expected using the nominal \swift-XRT effective area curve and an
infinite radius aperture ($R_{\mathrm{fake},0}$).  The 
count-rate measured in the $i$th visit, $R_{i}$, can then be detrended using the following relation
$R_{\mathrm{corrected},i} = R_{i} ( R_{\mathrm{fake},0} / R_{\mathrm{fake},i})$.  
We assumed a simple spectral model consisting
of a power-law slope of $\Gamma=1.9$ and Galactic $N_H =
1.2\times10^{22}$~cm$^{-2}$, but note that the detrending method is
only weakly dependent on spectral shape due to our use of medium-width
energy bands.

The detrended count-rate in the 3--10~keV energy band was calculated by
summing the detrended count-rates calculated for the 3--4, 4--5, 
5--6, 6--7 and 7--10~keV bands. Uncertainties were calculated by summing in
quadrature the uncertainties in each band. 

In a number of visits the source position was located close 
to one or more of the bad columns on the \swift-XRT CCD, 
these data points have very uncertain correction factors and so are 
rejected from further consideration. Note that the \swift-XRT observations 
are expected to suffer from moderate pile-up effects when \axj\ is in its highest luminosity
state. However, as we do not carry out a spectral analysis of the \swift-XRT data
we do not attempt to correct for pile-up in this work.

\section{Accretion state determination} 
\label{State}

Figure \ref{LC} shows the long-term 3--10~keV light curve of \axj\ as measured by 
\swift-XRT. The locations and equivalent countrates for all \xmm\ observations plus 
those for the hardest and softest \nustar\ observations are also 
indicated\footnote{We first derive the 3--10~keV flux from the best-fitting model (see 
\S~4) for each of the \xmm\ and \nustar\ spectra, and then  
use WebPIMMS {\sc http://heasarc.gsfc.nasa.gov/Tools/w3pimms.html} to convert this to 
an equivalent \swift-XRT count rate.}.  
During 11/38 \xmm\ observations \axj\ was caught in outburst, with 
F$_{3-10\mathrm{keV}}\geq 1.6\times10^{-11}$\ergcms.
\axj\ was in quiescence during 19/38 \xmm\ observations, 
i.e. it was undetected in the 3--10~keV band, implying a luminosity 
lower than $\sim9\times10^{32}$~\ergps\ (assuming a distance of 8~kpc).
During the remaining 8/38 observations \axj\ was detected (see Tab. \ref{data}) 
at a slightly higher flux of 
F$_{3-10\mathrm{keV}}\leq1.3\times10^{-12}$~\ergcms\ corresponding to 
luminosities of L$_{3-10\mathrm{keV}}\leq10^{34}$~\ergps, and Eddington ratios 
of $L_{3-10\mathrm{keV}}/L_{Edd} \leq5\times10^{-5}$, 
therefore, although detected, it has still been caught in the quiescent state. 

The observed high column density of neutral material 
($N_H\sim1.9\times10^{23}$~cm$^{-2}$) suggests that \axj\ 
is located at or behind the Galactic center (see Section \ref{NHs}). 
The distance of \axj\ along the line of sight is uncertain. 
If located at a distance of $\sim8-20$~kpc, \axj\ has a peak luminosity of 
$L\sim3-30\times10^{36}$~\ergps\ (see Section \ref{cont}), corresponding 
to several per cent of the Eddington luminosity (for a NS mass of $1.4$~M$_{\odot}$). 
The small angle between \axj\ and Sgr~A$^{\star}$, and the significant 
increase in surface density of stars and X-ray binaries towards the Galactic 
center (e.g. Muno et al. 2003), makes the Galactic center the most probable 
location. Therefore we assume a distance to \axj\ of 8~kpc. 
Such a distance is further supported by observations of the brightest X-ray bursts
emitted by \axj, which reached the Eddington luminosity for a NS 
at 8~kpc (Maeda et al. 2006, Degenaar et al. 2009). The luminosities 
computed in this work can be scaled by a factor 
$(\mathrm{dist}/8~\mathrm{kpc})^2$ should a more reliable measurement 
of the distance of \axj\ become available. 

It is well known that, at luminosities in the range $1-30$~\%~$L_{\rm Edd}$, 
NS-LMXBs typically alternate between two distinct states (e.g. van der Klis 2006), 
with state transitions following a hysteresis pattern in the Hardness Intensity 
Diagram (Mu\~{n}oz-Darias et al. 2014). 
The X-ray spectra of NS-LMXBs in outburst (see Lin et al. 
2007; Barret et al. 2001) are characterised by two main states: 
i) a hard state where the 3--10~keV emission is dominated 
by a power-law component 
and strong variability ($rms$ up to $\sim$20--40~\%); 
ii) a soft state where the X-ray 
emission is dominated by a disc blackbody component 
and there is only weak broad-band variability ($rms<5$~\%). 
Note that the same levels of rms and spectral properties 
characterize the hard and soft state in black hole X-ray binaries 
as well (e.g. Mu\~{n}oz-Darias et al. 2011). 
NS-LMXBs can also show an additional X-ray emission component, 
associated with emission from the surface of the neutron star. 
Furthermore, the observed spectrum of \axj\ is significantly modified 
by absorption from a high column density of neutral material 
(consistent with observations of other Galactic center sources). 

To determine the state of \axj\ in outburst, we investigated two independent 
measures: the X-ray variability and the shape of the X-ray continuum. 
We fitted the spectrum from each \xmm\ {\sc obsid} with three different continuum models 
(all absorbed by a column density of neutral material fitted with the \texttt{phabs} model 
with Anders \& Grevesse 1989 abundances and Balucinska-Church \& McCammon 
1998 cross sections).
They are: i) a multi-temperature disc blackbody (\texttt{phabs$\times$diskbb}); 
ii) a single temperature blackbody component (\texttt{phabs$\times$bbody}); and iii) a power-law 
(\texttt{phabs$\times$powerlaw}) continuum model. 
We report in Table \ref{fit} the $\chi^2$ and best fit values for the power-law model 
and disc blackbody as well as the differences between the $\chi^2$ for the  
best-fitting power-law model compared to the $\chi^2$ for the best-fitting blackbody model. 
We find that for each of the observations where \axj\ has a high flux, the thermal 
(blackbody and disc blackbody) models give a significantly better fit compared to 
the power-law model. The power-law model is preferred in the two observations 
at the lowest flux. At the highest fluxes the best-fitting power-law spectral index 
indeed assumes very steep values ($\Gamma\sim2.8-3.1$) suggesting instead the presence 
of a thermal component. Despite the blackbody and disc blackbody models 
well reproducing the shape of the continuum of the high flux observations, the presence 
of very significant residuals in the Fe~K band make the fit formally unacceptable 
(see Tab. \ref{fit}). We note that an acceptable fit is obtained once the Fe~K band 
($5.5<E<8.5$~keV) is excluded (see column 9 in Tab. \ref{fit}).

We also investigated the X-ray variability of \axj. We produced light curves in the 
3--10~keV energy band\footnote{We consider only the hard X-ray band because, 
due to the high column density of neutral material towards \axj, only a small fraction 
of $E<3$~keV source photons reach us (see e.g. Maeda et al. 2006; Degenaar et al. 
2010a; or Fig. \ref{lddel}).} from the \xmm\ EPIC-pn data, 
with 73~ms time resolution\footnote{This is the finest time resolution available for the PrimeFullWindow 
observing mode.}, and cleaned of time intervals associated with dips and bursts 
(using the same criteria and thresholds used in Section~2.1 and Table~\ref{data}). 
These light curves were used to compute the power spectral density (PSD) 
function of each {\sc obsid}, and to derive an estimate of the 
high frequency (i.e. freq~0.1--7~Hz) fractional $rms$. However, since the detected 
EPIC-pn count rate is typically fairly low (i.e. between ~0.8 and 7~ct~s$^{-1}$) the PSDs are 
dominated by counting noise and do not permit a reliable measurement of the intrinsic 
source variability. 

\begin{figure}
\includegraphics[width=0.35\textwidth,angle=-90]{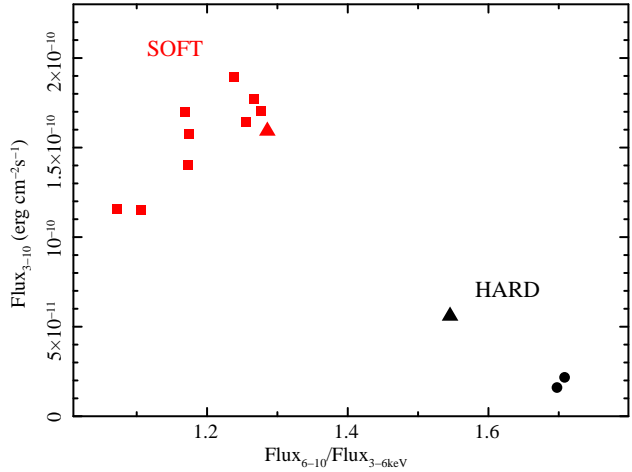}
\caption{Hardness Intensity Diagram for \axj. Red squares and black circles 
indicate soft and hard state \xmm\ observations, respectively. Soft state 
observations are fitted best by a simple blackbody or disc blackbody model, 
while hard states are fitted best by a power-law emission model. Triangles show the 
hardest and softest \nustar\ observations used to constrain the source 
SED (to compute proper ionised absorber models, Section \ref{SEDmod}).
}
\label{HID}
\end{figure}

Finally, we identify the position of each observation of \axj\ in the Hardness 
Intensity Diagram (HID; see Fig. \ref{HID}), which is often used to determine the source 
state (Fender et al. 2004; Belloni et al. 2011; Mu\~{n}oz-Darias et al. 2014). 
The hardness-ratio is defined here as the ratio between the 
observed fluxes in the 6--10~keV and the 3--6~keV bands. As expected, thermal emission 
dominated observations have a markedly smaller hardness-ratio compared to 
power-law dominated ones. 
Note that the observed hardness-ratio indicates a real variation of the SED (e.g. it is not 
due to a variation of the neutral absorber; see Section \ref{NHs}).
We therefore denote the thermal emission softer observations as `soft state', 
and the power-law dominated harder observations as `hard state'.

\section{Phenomenological models} 

In what follows (guided by the results presented in Section \ref{State}), we consider only an absorbed 
power-law spectral model for the hard state observations, and either an absorbed disc blackbody 
or an absorbed single blackbody component for the soft state observations (see Tab. \ref{fit}). 

\subsection{Neutral absorbing material}
\label{NHs} 

\begin{figure}
\includegraphics[width=0.34\textwidth,angle=-90]{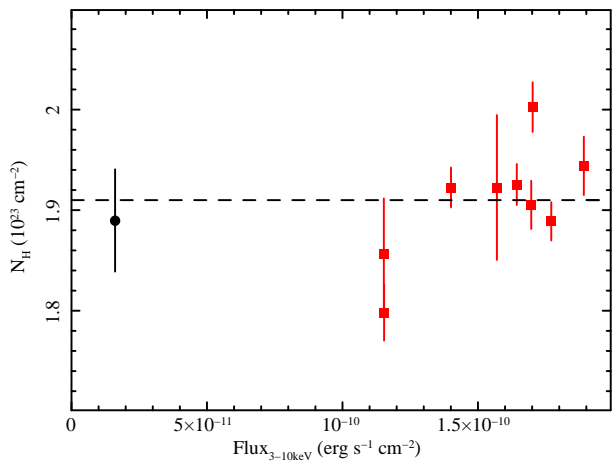}
\caption{Column density (in units of $10^{23}$~cm$^{-2}$) 
of neutral material as a function of the observed 3--10~keV flux. Red squares 
and black circles show soft- and hard state observations, respectively. }
\label{NH}
\end{figure}
Fig. \ref{NH} shows the best-fitting column density 
of neutral material during the soft and hard state observations. 
These measurements are consistent with a constant column density, despite the very large variation 
in source flux and spectral shape. This suggests that, despite \axj\ being a dipping 
source, the observed column density of neutral material is most probably attributable  
to absorption along the line of sight, i.e. physically unrelated to \axj. 
In fact, the observed column density is within a factor~2 of the column densities measured
towards two other nearby Galactic centre sources: \sgras\ and SGR~J1745-2900, 
$N_H=1.1$ and $1.7\times10^{23}$~cm$^{-2}$ respectively 
(Baganoff et al. 2003; Trap et al. 2011; Nowak et al. 2012; Rea et al. 2013). 
This high column density of neutral material supports the hypothesis that \axj\ is located 
either at or behind the Galactic center. 

\subsection{Thermal emission in the soft state}
\label{cont}

\begin{figure}
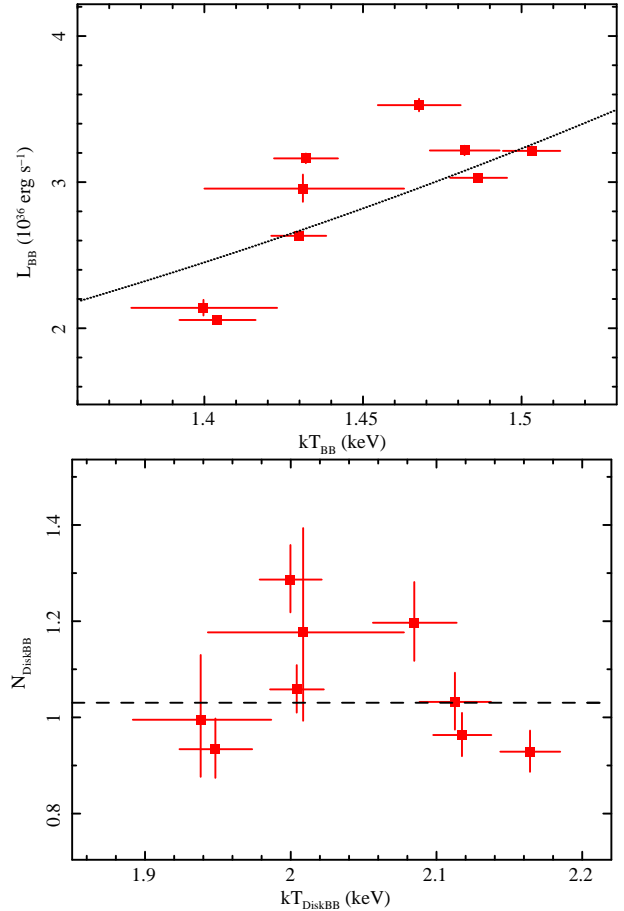

\includegraphics[width=0.34\textwidth,angle=-90]{BBNH2.ps}
\includegraphics[width=0.34\textwidth,angle=-90]{DBBNH2.ps}
\caption{{\it (Top panel)} Observed disc luminosities (in units of $10^{36}$~erg 
s$^{1}$) versus disc temperature (in keV) obtained from fits with the 
\texttt{phabs$\times$bbody} model. The doted line shows the best 
fit relation (L$_{BB}\propto kT_{BB}^4$). 
{\it (Bottom panel)} Multi-temperature disc blackbody normalisation as 
a function of disc temperature. }
\label{BBNH}
\end{figure}
The upper panel of Fig. \ref{BBNH} shows the best-fitting temperature and
luminosity for the soft state spectra fitted with a simple blackbody.
During these observations \axj\ is observed at a luminosity between
$L\sim2-4\times10^{36}$~\ergps. Given the high-inclination of
the system, it is probable that the intrinsic disc luminosity is much
higher. For example, simple geometrical considerations (not taking
into account relativistic effects) suggest that for \axj's
inclination angle of $\theta\sim80^\circ$, the intrinsic
disc luminosity should be a factor of
$cos(\theta\sim80^\circ)^{-1}\sim3$ higher than the observed
luminosity\footnote{The inclination of \axj\ was proposed to be $\sim 70^{\circ}$ 
by Maeda et al. (1996). However, this value is a lower limit since (i) it assumes 
a 1~$M_{\odot}$ main sequence companion and (ii) the eclipses should be grazing. 
On the other hand, an inclination $\gsimeq 85^{\circ}$ is not expected since the 
system does not show the typical properties of accretion disc corona sources 
(e.g. White \& Holt 1982). Since the companion stars in LMXBs are typically 
evolved (and thus less massive than expected for a main sequence star), and 
this source displays relatively deep eclipses, we adopt an inclination angle 
of $80^{\circ}$. In any case, we note that our results are not 
significantly dependent on small variations of this orbital parameter.}. 
Therefore the intrinsic disc luminosity of \axj\
corresponds to $\sim4-10$~per~cent Eddington (versus 
$\sim1.2-2.6$~per~cent Eddington for the uncorrected luminosity).  
This places \axj\ in the typical luminosity range observed in `atoll' 
sources (Gladstone et al. 2007; Mu\~{n}oz-Darias et al. 2014). 
\axj\ would be characterised by even higher luminosities if it is in fact 
located far behind the Galactic center.

The upper panel of Fig. \ref{BBNH} shows a clear trend of larger luminosities 
with increasing blackbody temperatures. Blackbody temperature 
and luminosities are well correlated with a Spearman correlation coefficient 
of 0.78 (with an associated null hypothesis probability of $1.3\times10^{-2}$
corresponding to a confidence level of $98.7$~per~cent)\footnote{
See Appendix A of Bianchi et al. (2009; see also Ponti et al. 2012b) for details 
on this procedure, which takes into account errors on the $Y$ variable.}. 
The line in Fig. \ref{BBNH} shows the relation, $L_{BB}\propto kT_{BB}^4$ 
that is expected when the thermally emitting region varies in temperature but keeps 
a constant area. This simple relation reproduces the observed trend, however significant 
intrinsic scatter is present. This scatter might be produced by the emission from 
the boundary layer. 
In addition, the $L_{BB}\propto kT_{BB}^4$ trend is supported by the results of spectral fits using 
a disc blackbody component (see lower panel of Fig. \ref{BBNH}), which, to first  
approximation, are consistent with a disc having constant normalisation and an 
inner radius $R_{in}\sim12$~km (assuming a disc inclination of $80^\circ$ and 
a color to effective temperature ratio $k=2$; Shimura \& Takahara 1995; Kubota et al. 1998). 

\subsection{Soft and hard state comparison}

\begin{figure*}
\includegraphics[width=0.8\textwidth]{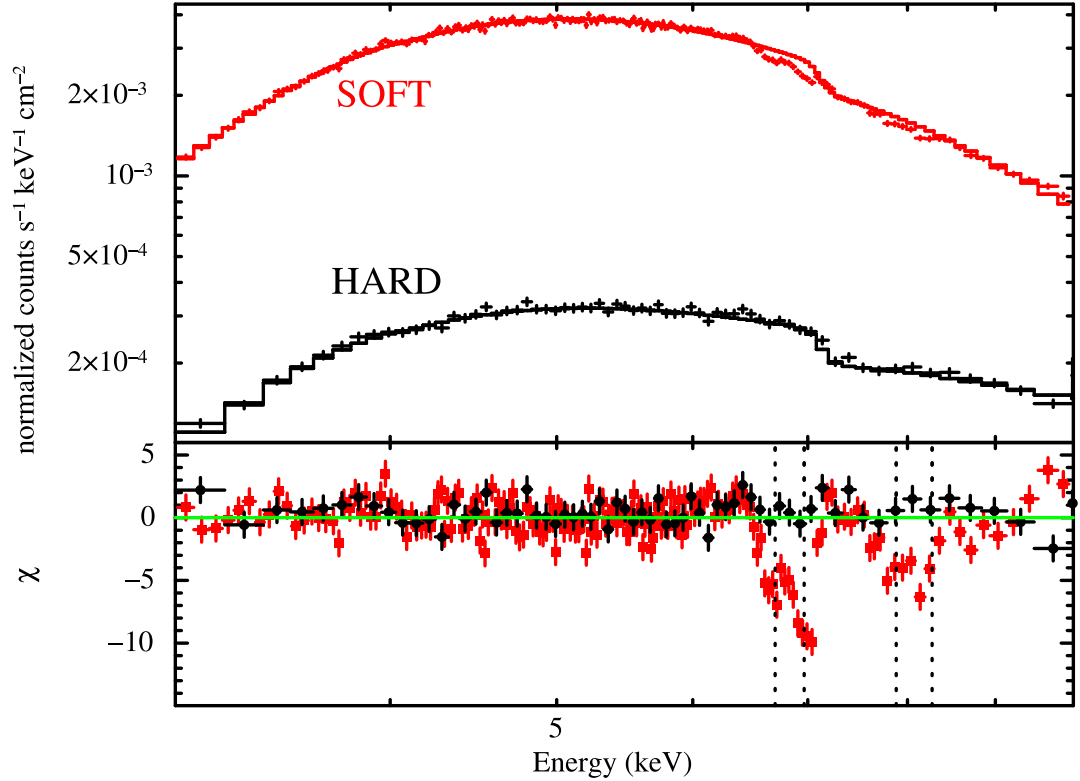}
\caption{The combined spectrum from all soft state \xmm\ observations is shown 
with red squares. 
The combined spectrum for hard state observations is shown with black circles. 
The continuum is fitted with an absorbed 
blackbody (soft state) and power-law (hard state), see Section \ref{cont}. 
Significant absorption features at the energies of the \Fevc~K$\alpha$ 
and \Fevs~K$\alpha$, as well as \Fevc~K$\beta$ and \Fevs~K$\beta$ (indicated 
by the dotted lines) are clearly visible in the combined spectrum for the soft state observations. 
There are no obvious residuals due to ionised narrow absorption lines in the 
combined spectrum for the hard state observations.}
\label{lddel}
\end{figure*}
Fig. \ref{lddel} shows the combined time-averaged spectra of all the soft- and 
all the hard state \xmm\ observations, and their residuals with respect to 
an absorbed blackbody model (soft state) and to an absorbed power-law model (hard state).

Systematic negative residuals at the energies of the \Fevc\ and \Fevs\ transitions 
are clearly visible in the time-averaged soft state spectrum.
These absorption features are also visible, albeit at lower signal to noise ratio, 
in the spectra from individual soft state observations.
The combined soft state spectrum shows very significant residuals at the energies of the 
\Fevc~K$\alpha$ ($E=6.697$~keV) and \Fevs~K$\alpha$ ($E=6.966$~keV), as well 
as \Fevc~K$\beta$ ($E=7.880$~keV) and \Fevs~K$\beta$ ($E=8.268$~keV) lines,  
with a possible contribution from Ni~{\sc xxvii}~K$\alpha$ and Fe~{\sc xxv}~K$\gamma$ lines.
Similar absorption lines were also observed by Hyodo et al. (2009) in a single \suzaku\ 
observation.  
These features are therefore most probably produced by absorption from photo-ionised gas. 
In contrast, no significant narrow negative residuals are visible in the time averaged hard 
state spectrum, nor in the spectra from the individual hard state observations, with 
upper limits to the absolute value of the line equivalent width as stringent 
as $\sim5-15$~eV.

This one-to-one correlation of wind-absorption with accretion state appears 
similar to that seen in the other well monitored neutron stars, \exo\ (Ponti et al. 2014), 
and in Galactic black hole binaries (Miller et al. 2008; 2012; Neilsen \& Lee 2009; 
Ponti et al. 2012). 

\subsection{\Fevc\ and \Fevs\ absorption line inter-observation variability}

In order to examine in more detail the variability of these absorption lines, we fit the 
spectrum from each 
soft state observation with an absorbed single blackbody model plus two narrow absorption lines: 
(\texttt{phabs$\times$(gaus+gaus+bbody)}). 
Such lines have been observed in most high-inclination low-mass X-ray binaries and 
they are typically unresolved at the EPIC-pn CCD resolution (Diaz-Trigo et al. 2006). 
Hoever, grating observations of 
such features reveal typical widths of $\sim$~few~$10^2$~km~s$^{-1}$ 
for these lines (Miller et al. 2008; Kallman et al. 2009). 
Therefore, at first, we fix the line widths of the \Fevc\ and \Fevs\ 
absorption lines to 5~eV. For the measurement of absorption lines,
we consider only the seven soft state \xmm\
observations having clean exposure times longer than 5~ks (see Tab. \ref{fitlines}). 
The addition of the two narrow absorption lines significantly improves the fit 
(compared to a simple absorbed blackbody model) from 
$\chi^2=3333$ for 2324 dof to $\chi^2=2578$ for 2296 dof. 
Clear residuals are still observed at energies corresponding to the Fe~{\sc xxv}~K$\beta$ 
and Fe~{\sc xxvi}~K$\beta$ lines. Addition of these two ionised narrow 
(with width fixed to 5~eV) Fe~K$\beta$ absorption lines to the spectral model 
significantly improved the fit to $\chi^2=2334$ for 2270 dof. 
The best fit parameters of these fits are reported in Tab. \ref{fitlines}. 
To test whether or not these lines appear resolved, we then tie the width of all 
the lines to a single value that is left free to vary. 
We obtain a best-fitting energy of $\sigma=0.035\pm0.015$~keV, corresponding 
to $v\sim1400\pm500$~km~s$^{-1}$. However, we observe only a small improvement 
of the fit ($\chi^2=2328$ for 2269 dof, with associated F-test probability $0.018$), 
suggesting that these absorption lines are not (or are barely) resolved. 
Therefore, in the analysis that follows we keep the line widths fixed to 5~eV. 

\subsubsection{Fe~K line energies}

\begin{figure*}
\includegraphics[width=0.33\textwidth]{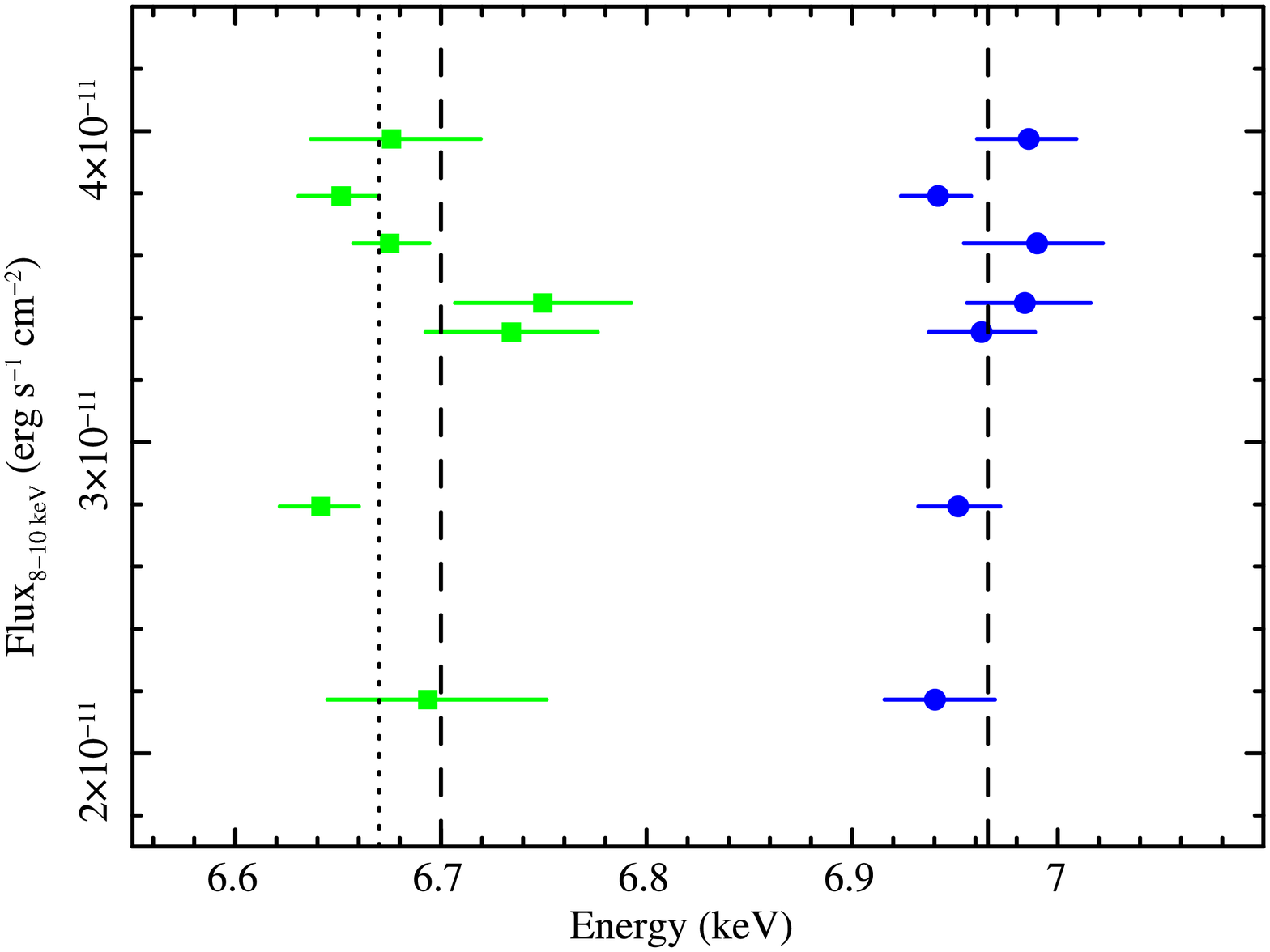}
\includegraphics[width=0.33\textwidth]{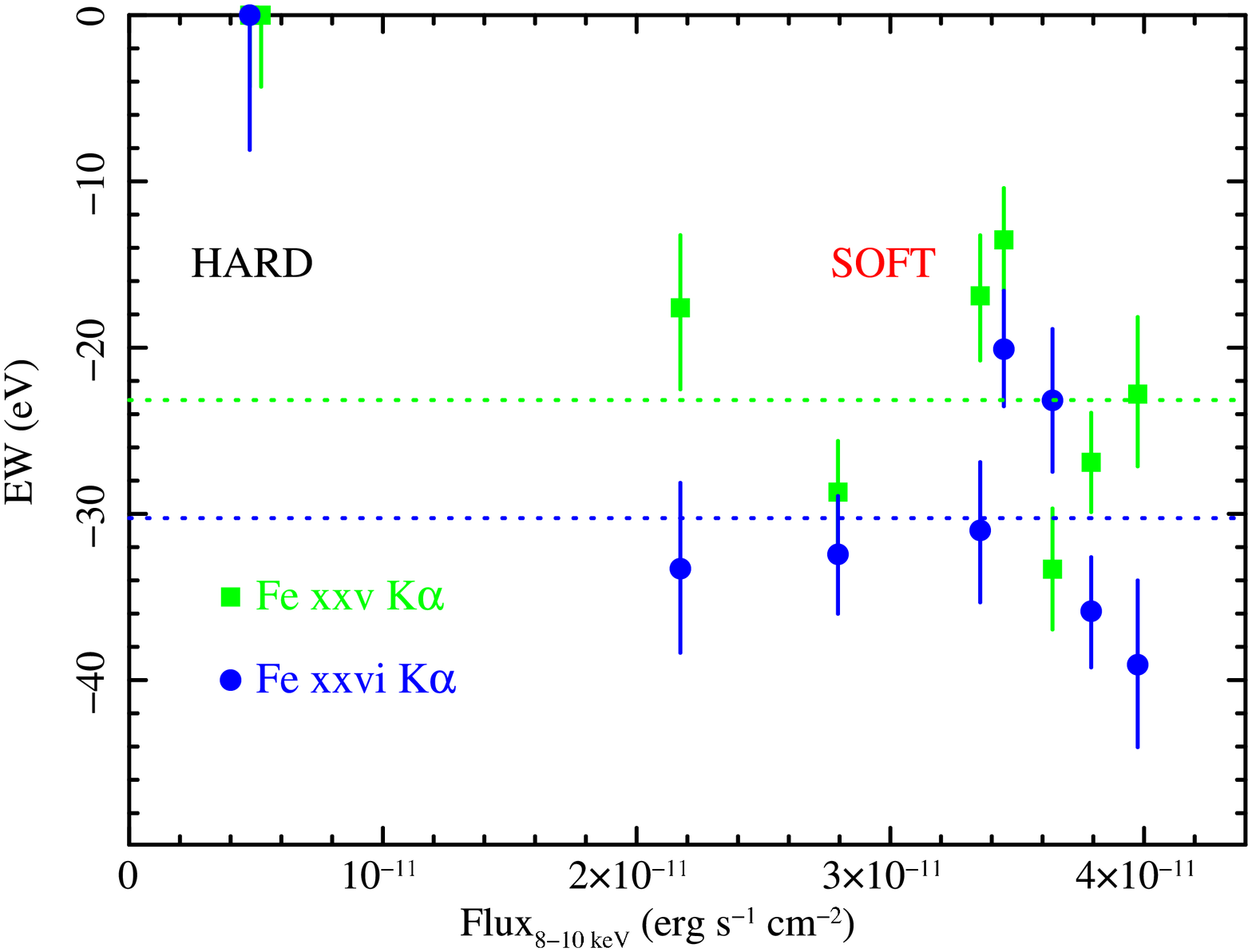}
\includegraphics[width=0.33\textwidth]{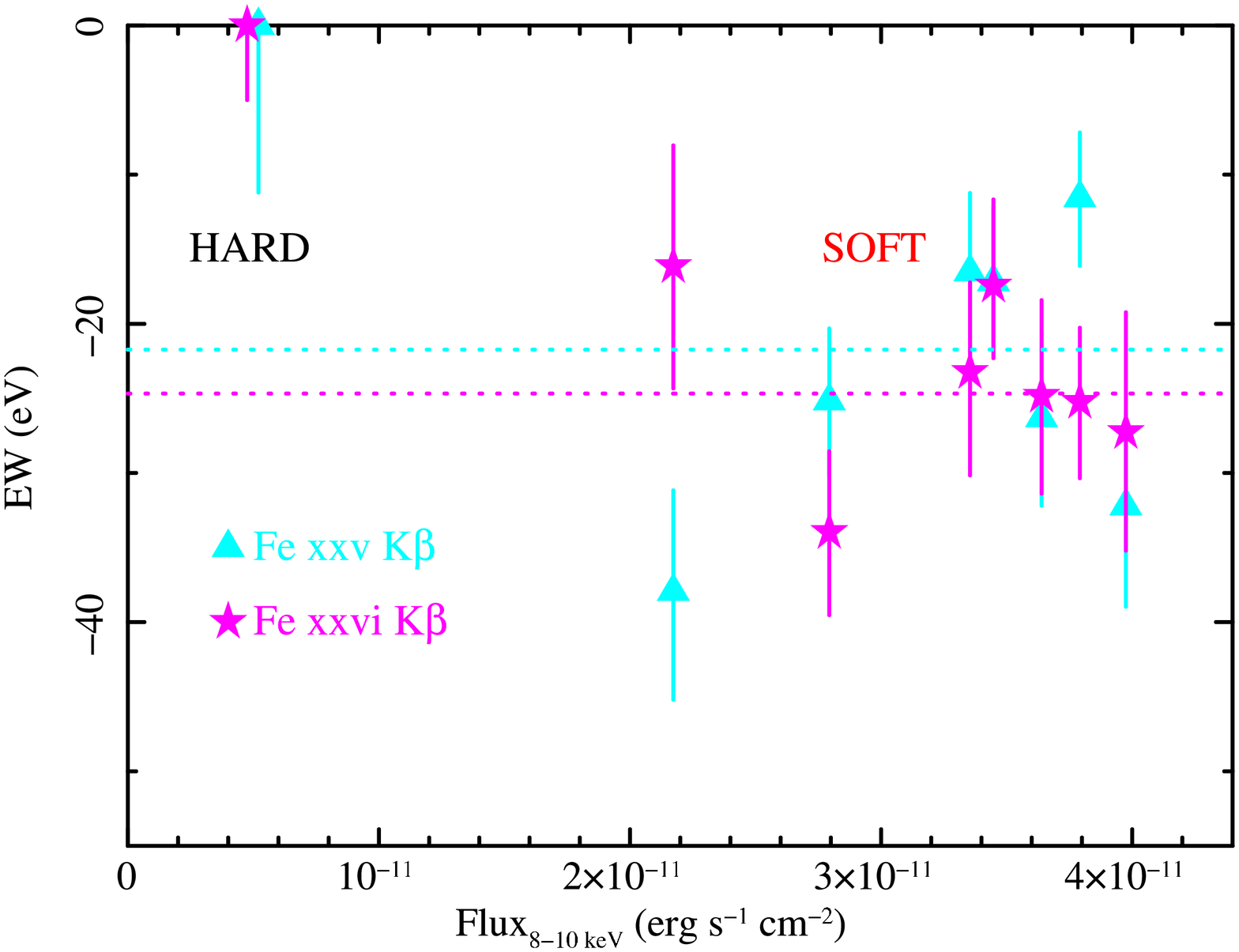}
\caption{{\it (Left panel)} Best-fitting energies of the \Fevc\ and \Fevs~K$\alpha$ lines. 
The vertical lines indicate the energies of the resonance
\Fevs\ and \Fevc\ lines (dashed) and the inter-combination \Fevc\ transition (dotted). 
{\it (Middle panel)} Equivalent widths of the \Fevc\ and \Fevs\ K$\alpha$ 
absorption lines (shown in green and blue, respectively) as a function 
of the 8--10~keV flux. Both lines are significantly detected in each soft 
state observations, and stringent upper limits can be placed on the lines during the  
hard state observation. 
{\it (Right panel)} Equivalent widths of the \Fevc\ and \Fevs\ K$\beta$ absorption 
lines (shown in cyan and magenta, respectively) as a function of the 8--10~keV flux. 
In all three panels the error bars indicate the one sigma uncertainties (or limits).}
\label{LinesE}
\end{figure*}
The left panel of Fig. \ref{LinesE} shows the best-fitting energies of the
\Fevc~K$\alpha$ and \Fevs~K$\alpha$ absorption features and the expected 
energies of the resonance \Fevs~K$\alpha$ and \Fevc~K$\alpha$ lines and 
inter-combination \Fevc~K$\alpha$ transitions (see also Tab. \ref{fitlines}). 
No significant blue- or redshift is observed. 
However, due to the finite energy resolution and the uncertainty 
on the energy scale calibration of the EPIC-pn camera, only outflows with
$v_{out}\gg10^3$~km~s$^{-1}$ would be detected.  Higher energy
resolution observations (together with better energy scale
calibration) are thus required to measure outflows that are typically
observed in Galactic BHs ($v_{out}\sim10^2-10^3$~km~s$^{-1}$).  Note
that the \Fevc\ and \Fevs\ K$\beta$ absorption features are observed
at the expected energies of these transitions (see Tab. \ref{fitlines}).

\subsubsection{Fe~K line equivalent widths}

The middle and right panels of Fig. \ref{LinesE} show the observed 
equivalent widths of the \Fevc, \Fevs\ K$\alpha$ and K$\beta$ absorption 
lines (see also Tab. \ref{fitlines}). 
These four absorption lines are all significantly detected in the 
individual spectrum from each soft state observation,
with average equivalent widths in the range $\sim-40$ to $-15$~eV. 
In contrast, from the single hard state spectrum we can only derive stringent upper 
limits (see Fig. \ref{LinesE} and Tab. \ref{fitlines}). 
Within the soft state observations, we observe no clear trend of increasing or decreasing 
line equivalent widths with 8--10~keV flux.  

\begin{figure*}
\includegraphics[width=0.43\textwidth]{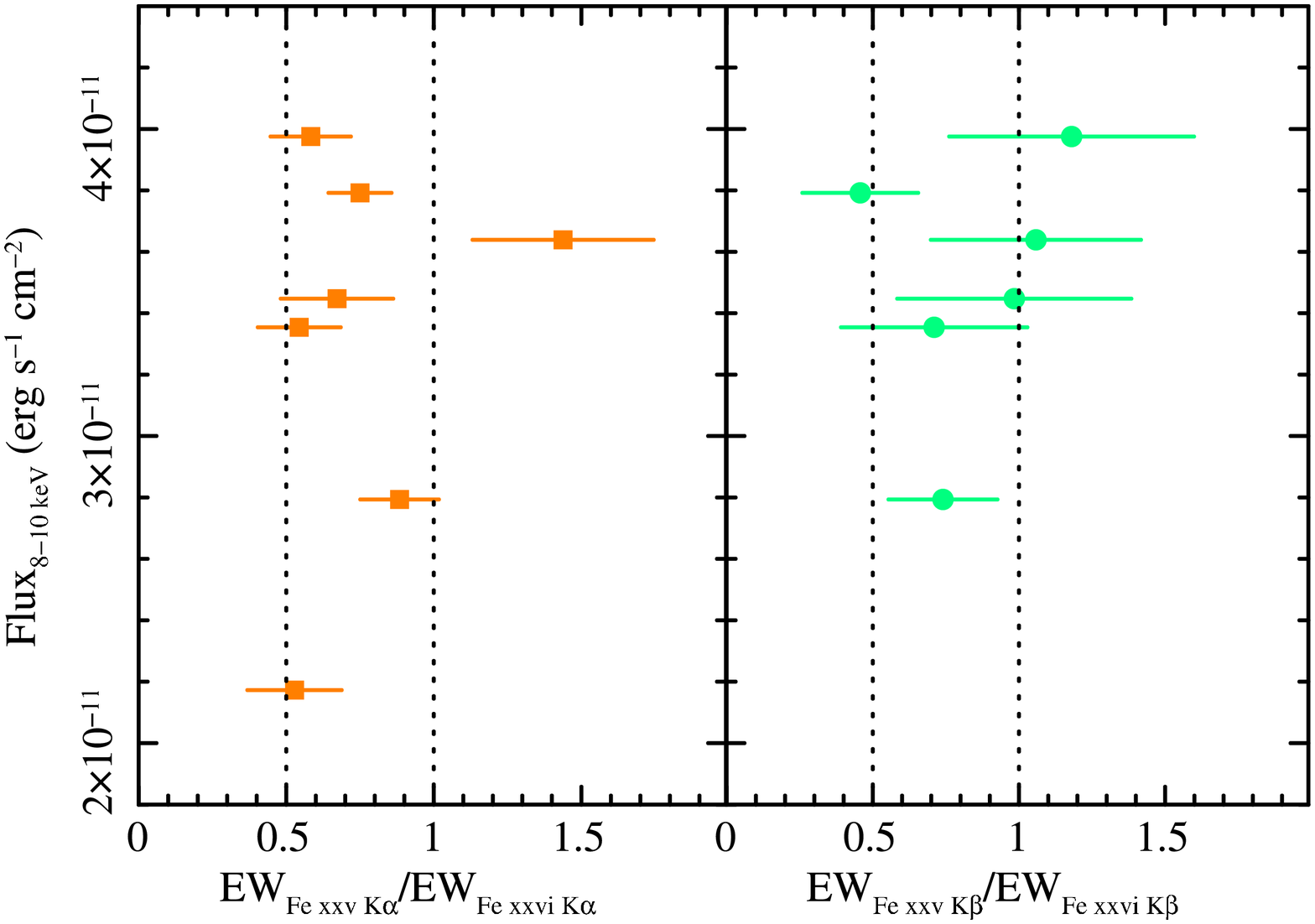}
\includegraphics[width=0.43\textwidth]{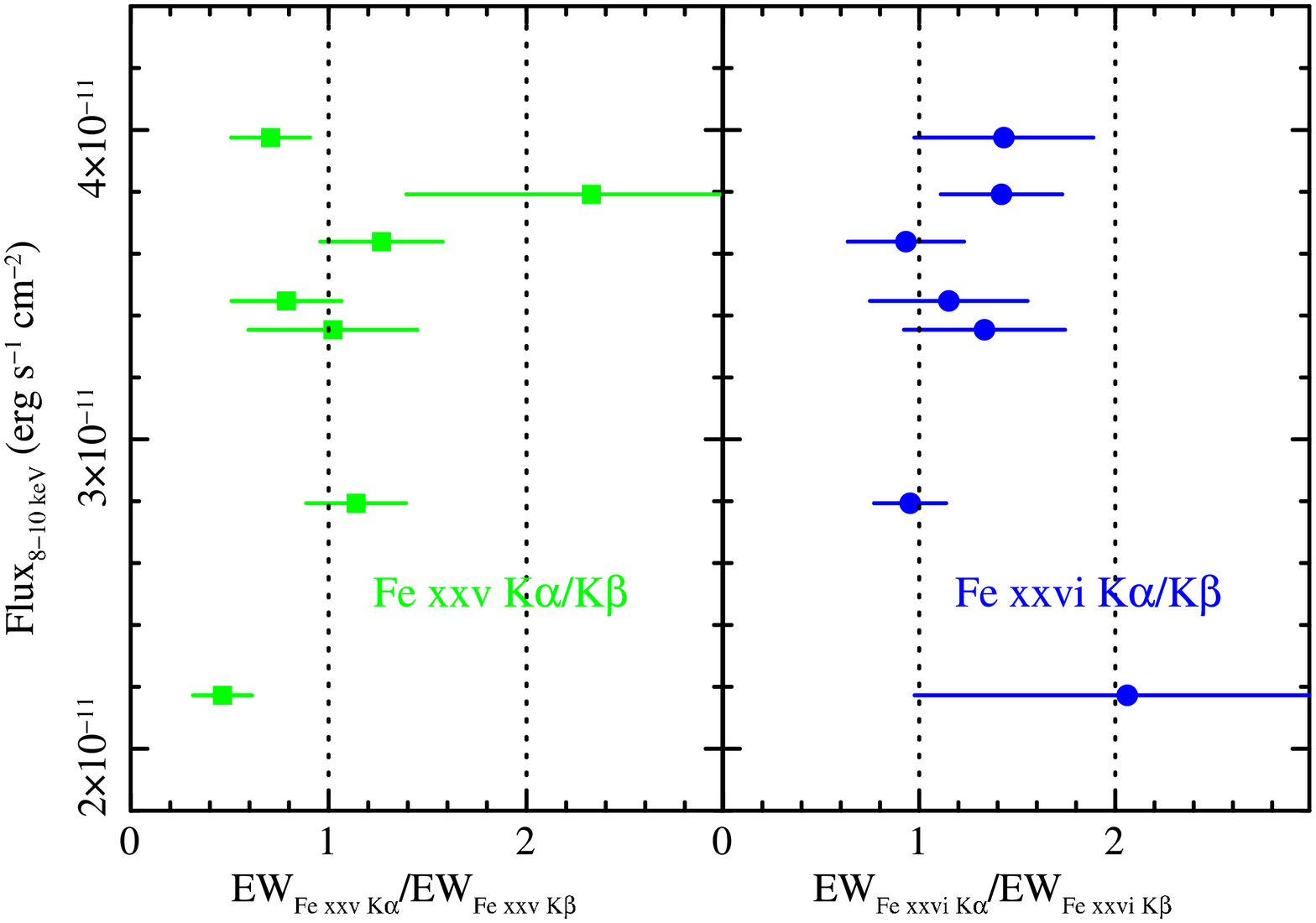}
\caption{{\it (Left panel, left half)} Ratios of the equivalent width of the \Fevc\ K$\alpha$ line to  
the equivalent width of the \Fevs\ K$\alpha$ line as a function of the 8-10~keV 
flux. 
{\it (Left panel, right half)} Same for the \Fevc\ K$\beta$ and \Fevs\ K$\beta$ lines. 
{\it (Right panel, left half)} Same for the \Fevc\ K$\alpha$ and \Fevc\ K$\beta$ lines. 
{\it (Right panel, right half)} Same for the \Fevs\ K$\alpha$ and \Fevs\ K$\beta$ lines. 
Error bars show the 1~$\sigma$ uncertainties. }
\label{LinesEWRat}
\end{figure*}
Figure \ref{LinesE} shows that both the \Fevs\ K$\alpha$ (middle panel) 
and K$\beta$ (right panel) lines are typically more intense ($\sim15-20$~\%) than the 
corresponding \Fevc\ lines. 
The ratio between the equivalent widths of the \Fevc\ and \Fevs\ lines is 
a sensitive probe of variations in the ionisation state of the absorbing highly 
ionised plasma. The left panel of Fig. \ref{LinesEWRat} shows the \Fevc\ 
over \Fevs\ equivalent width ratios for both the K$\alpha$ and K$\beta$ lines. 
A small scatter between the different soft state observations is observed 
(at least for the K$\alpha$ transitions), suggesting that despite ionisation effects 
being present\footnote{Given the high densities expected within such ionised 
absorbers in binaries, the recombination time is expected to be much shorter 
than the typical time intervals between the \xmm\ observations.}, they do not 
play a major role here. In fact, no clear trend with flux is observed. 
This is probably due to the significant, but small, variations of the 8--10~keV flux. 

We also note that the K$\alpha$ lines are less than a factor of 2 stronger 
than the corresponding K$\beta$ lines. In particular, the \Fevc\ K$\beta$ lines 
have equivalent widths comparable to the corresponding K$\alpha$ lines, 
for both \Fevc\ and \Fevs. 
As shown for the case of NGC~3516 (fig. 4 of Risaliti et al. 2005), these 
Fe K$\alpha$/Fe K$\beta$ ratios suggest high column densities for the 
absorbing medium ($N_{H}\gsimeq10^{23}$~cm$^{-2}$) 
and high turbulent velocities of the order of $v_{\rm turb}\sim500-1000$~km~s$^{-1}$. 
Note that the \Fevc\ and \Fevs\ K$\beta$ lines could have a contribution 
from Ni~{\sc xxvii}~K$\alpha$ and Fe~{\sc xxv}~K$\gamma$, respectively 
(see e.g., Hyodo et al. 2009). However these transitions have oscillator 
strengths significantly lower than the nearby Fe~K$\beta$ transitions, 
therefore little contribution is expected. Also note that the possible contribution 
due to these lines has been considered by Risaliti et al. (2005).

\section{Photo-ionisation models}

To obtain realistic measurements of the ionisation state and column density 
of the absorbing ionised plasma, we now fit the observed spectra with realistic 
photo-ionisation models. Therefore we substitute the multi-Gaussian components 
with a single photo-ionised component. In such an analysis the relative strengths 
of the various absorption features are tied together in a physical manner.
\begin{figure}
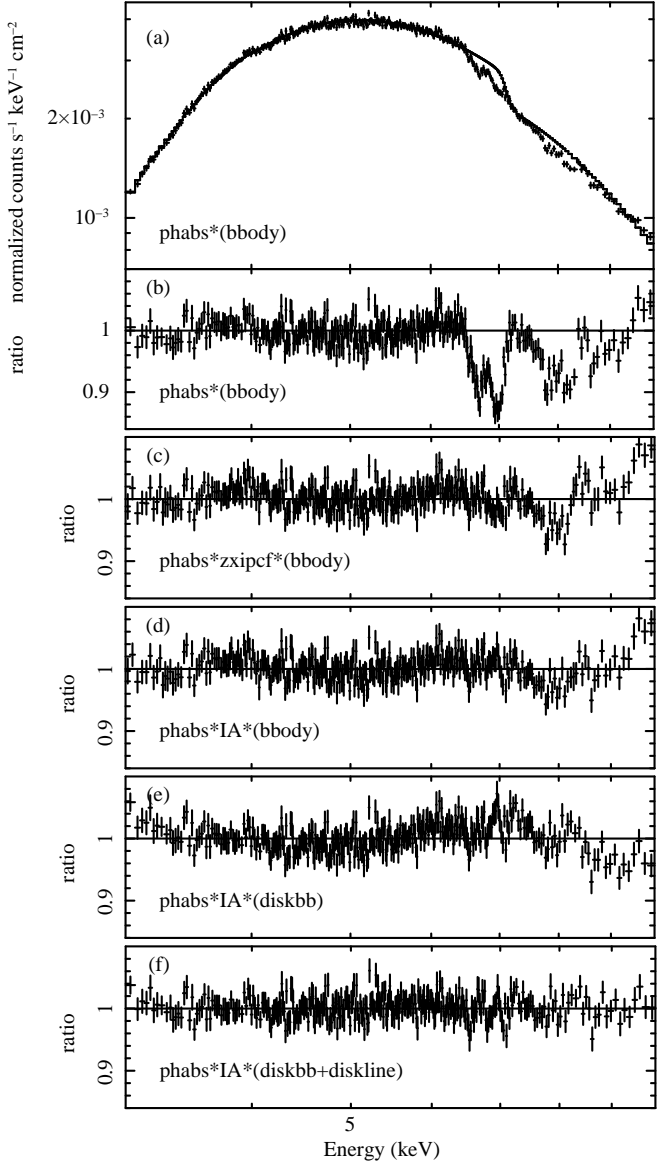

\includegraphics[width=0.37\textwidth,angle=-90]{zxipcf_soft500_NHfree_BB_ldel.ps}

\hspace{0.69cm}
\includegraphics[width=0.1255\textwidth,angle=-90]{zxipcf_del.ps}

\hspace{0.69cm}
\includegraphics[width=0.1255\textwidth,angle=-90]{zxipcf_soft500_NHfree_BB2.ps}

\hspace{0.69cm}
\includegraphics[width=0.1255\textwidth,angle=-90]{zxipcf_soft500_NHfree_diskBB2.ps}

\hspace{0.69cm}
\includegraphics[width=0.163\textwidth,angle=-90]{zxipcf_soft500_NHfree_diskBB_diskline2.ps}
\caption{{\it (Panel a):} Combined soft state spectra and best-fitting model (\texttt{phabs*IA*(bbody)}), 
once the Ionised Absorption (IA) component is removed. {\it (Panel b):} Same as panel (a), 
but showing the data to model ratio. {\it (Panel c):} Data to model ratio of the combined 
soft state spectra with the \texttt{zxipcf} component to fit the ionised absorption.  
{\it (Panels d, e and f):} Data to model ratio of the combined soft state spectra with the 
models: \texttt{phabs*IA*(bbody)}, \texttt{phabs*IA*(diskbb)} and \texttt{phabs*IA*(diskbb+diskline)}, 
respectively. }
\label{IAdel}
\end{figure}

\subsection{Fits with the {\it zxipcf} model}

\begin{figure}
\includegraphics[width=0.45\textwidth]{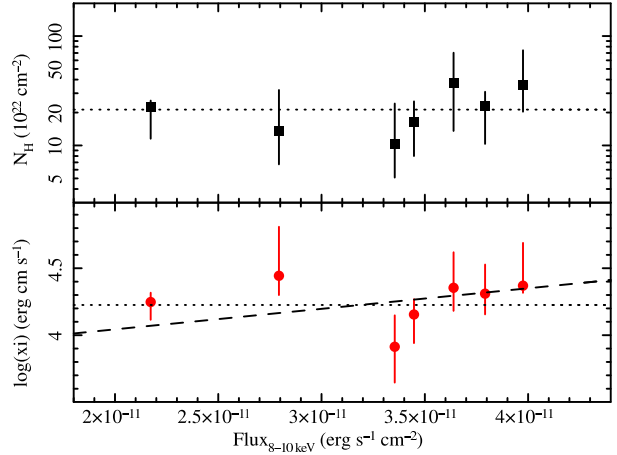}
\caption{Best fit column density ($N_H$, upper panel) and ionisation parameter 
($\xi$, lower panel) of the photo-ionised plasma, as measured using the \texttt{zxipcf} 
spectral model, as a function of the observed 8--10-keV flux. 
No significant variation is observed in either of the two parameters 
(the dotted lines show a model with constant $N_H$ and $\xi$). 
The dashed line shows the best-fitting linear relation between
ionisation parameter and source luminosity.}
\label{zxipcf}
\end{figure}
For an initial, approximate description of the ionised absorption, we use the \texttt{zxipcf} 
model, assuming that the obscuring plasma is totally covering the X-ray source: 
\texttt{phabs$\times$zxipcf$\times$(bbody)}. 
This component reproduces the absorption from photo-ionised gas 
illuminated by a power-law source with spectral index $\Gamma=2.2$ 
and is calculated assuming a micro-turbulent velocity of 200~km~s$^{-1}$ 
(Reeves et al. 2008; Miller et al. 2007). 
The addition of such a component drastically improves the fit 
($\Delta\chi^2=789.5$ for 14 new parameters) 
compared to the fit with a simple absorbed blackbody model (see Fig. \ref{IAdel}). 
We find best-fitting column densities for the photo-ionised plasma in the range 
$N_H\sim1-4\times10^{23}$~cm$^{-2}$ (see the upper panel of Fig. \ref{zxipcf}) 
and best-fitting ionisation parameters ${\rm log}({\xi/1{\rm ~erg~cm~s}^{-1}})=3.9-4.4$ 
(lower panel of Fig. \ref{zxipcf}). 
The column density of the ionised absorber is consistent with being constant between 
the different soft state observations (Fig. \ref{zxipcf}). 
Due to the high interstellar extinction towards \axj, we cannot detect additional 
absorption lines (besides Fe~K) and hence we can place only loose constraints on the 
plasma ionisation parameter. Furthermore the soft state \xmm\ observations span 
only a small range ($\sim$ a factor or 2) in luminosity. Therefore we cannot 
discriminate between a constant ionisation parameter (dotted lines in Fig. \ref{zxipcf}) 
and a scenario where the ionisation parameter varies linearly 
with the source luminosity (as expected in the over-simplified case of a constant 
SED; see dashed line in Fig. \ref{zxipcf}). 

As a general comment, we note that the \Fevc\ and \Fevs\ K$\alpha$ and K$\beta$ 
lines are typically well reproduced by the \texttt{zxipcf} component, suggesting that 
these lines are indeed produced by photo-ionised absorbing material.  
However, we note significant residuals in several of the individual soft state spectra. 
This might be related to the small micro-turbulent velocity used to compute 
the \texttt{zxipcf} table, compared to the high values suggested by the 
Fe K$\alpha$/Fe K$\beta$ line ratios. 
Indeed, the earlier modelling with independent Gaussian absorption lines 
resulted in significantly better fits, suggesting that the \texttt{zxipcf} component 
cannot fully reproduce all the details of the observed ionised absorption (see 
Fig. \ref{IAdel}). 

\subsection{Characteristic soft and hard state SEDs}
\label{SEDmod}

\begin{figure}
\includegraphics[width=0.48\textwidth,angle=0]{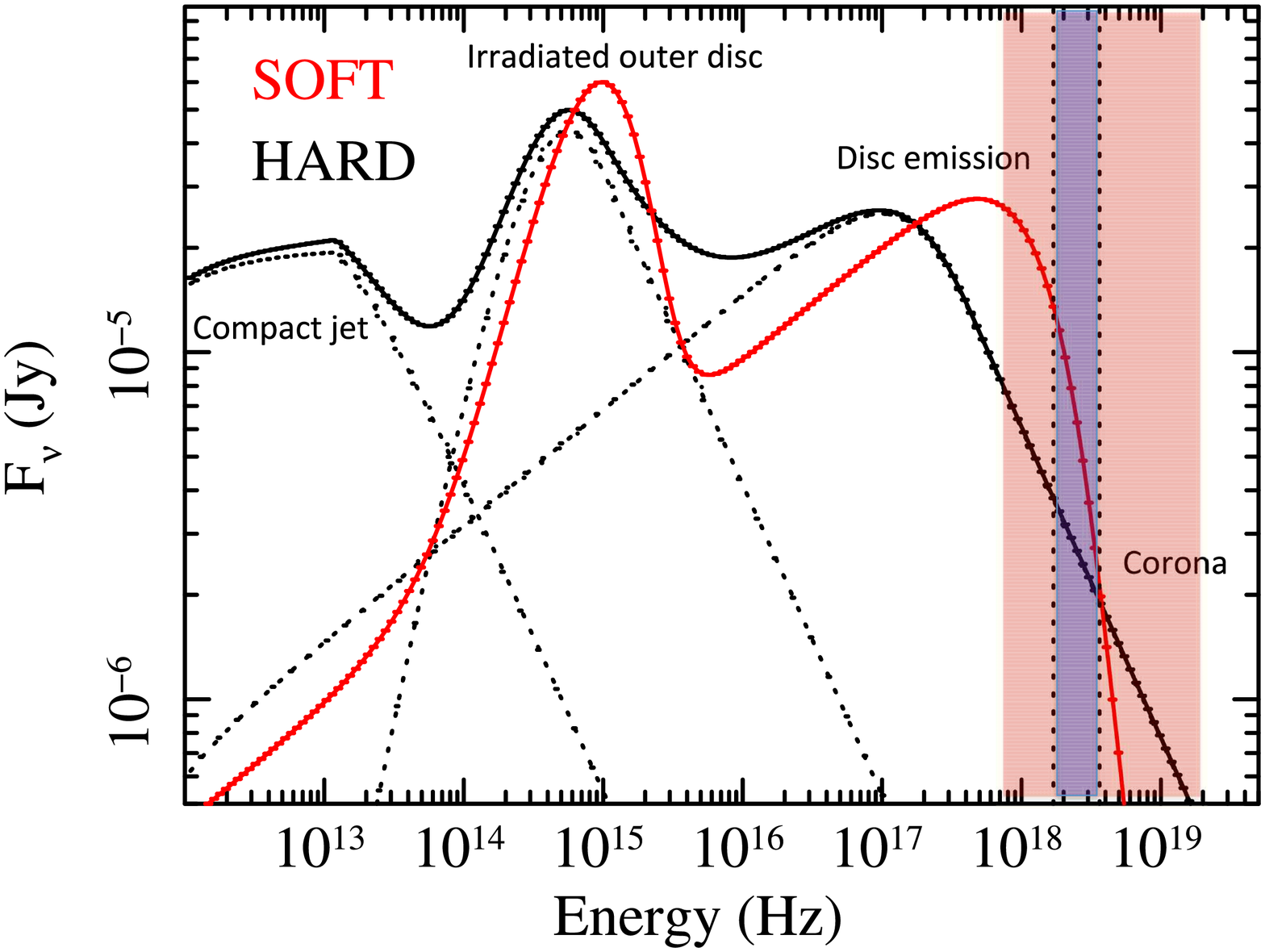}
\caption{Modelsfor the typical intrinsic SED of \axj\ during the 
soft (red line) and hard states (black line). 
The effects of intervening neutral absorption are removed for clarity. 
The key ingredients 
described in Section \ref{SEDmod} to produce the source SED are labelled. 
The shaded salmon-colour region indicates the energy band over which 
the SED is constrained by the \nustar\ observations ($E=3-79$~keV). 
The violet region highlights the energy band ($E=7-15$~keV) 
just above the Fe~K edge, most critical for producing Fe~K emission 
lines. }
\label{SED}
\end{figure}
To compute a physically consistent photoionisation model for the Fe~K absorption 
lines, the determination of the source flux and spectral shape in the $\sim7$ to 
$\sim10-20$~keV band is of primary importance (see the blue shaded band 
in Fig. \ref{SED}). Photons below $7.1$~keV can 
not ionise the Fe~K shell, therefore they can not produce either \Fevc\ or \Fevs\ 
absorption lines. However, the physical properties of the ionised absorber 
(e.g. the plasma temperature) do also depend, albeit weakly, on lower and higher 
energy photons. 
To determine the extension of the source emission above $\sim10$~keV, we fit the 
hardest and softest \nustar\ observations of the 2013 campaign to observe 
\sgras\ (Hailey et al. in prep.). 
We fit the \nustar\ spectra over the $3-79$~keV energy band (see the pink band 
in Fig. \ref{fitlines}) with 
an absorbed disc blackbody model providing the seed photons that are Comptonised 
and generate a power-law component (\texttt{phabs$\times$simpl$\times$diskbb}; 
Steiner et al. 2009). This allows us to constrain the emission from \axj\ in the full X-ray band. 
As typical for the soft state (Lin et al. 2007; Plant et al. 2014), the multi-temperature disc 
blackbody emission is well constrained and a very weak and steep power-law 
component is observed (see Hailey et al. in prep. for more details). 
A colder disc blackbody emission component 
is expected to be present during the hard state. 
Unfortunately, due to the significant Galactic absorption towards \axj, our data are not sensitive to 
the disc emission during the hard state. We, therefore assume a 
fixed temperature of $T_{disc}=0.3$~keV (e.g. Plant et al. 2014a,b).
We also check that the derived ratio between fluxes of the disc and power-law components are 
within the typical range observed during the soft and hard states (Lin et al. 2007; 
Remillard \& McClintock 2006). 
Optical and infrared observations of accreting 
black holes and neutron stars in the soft and hard states show evidence for irradiation of the outer disc 
(Hynes et al. 2002; Migliari et al. 2010). To model this, we added a thermal 
(blackbody) component with temperature $T_{BB-S}=15000$~K and 
$T_{BB-H}=7000$~K for the soft and hard state, respectively (Hynes et al. 2002). 
Finally, at radio-to-infrared frequencies we added a contribution from a 
compact jet, which is only observed in the hard state (Migliari et al. 2010). 
The soft and hard state SED are displayed with red and black lines in Fig. \ref{SED}, 
respectively. 

\subsection{Self-consistent photo-ionised models}

We prepared two ad-hoc tables (denoted IA$_{\rm{soft}}$ and 
IA$_{\rm{hard}}$) produced with the photo-ionization code Cloudy 
C13.00 (last described in Ferland et al. 2013). The model ingredients are: 
(1) the soft and hard SEDs presented in Sect. \ref{SEDmod}; 
(2) constant electron density $n_e = 10^{12}$ cm$^{-3}$; 
(3) ionization parameter in the range ${\rm log}(\xi/1{\rm ~erg~cm~s}^{-1}) = 3.0:5.0$; 
(4) intervening column density in the range ${\rm log}(N_H/1{\rm ~cm^{-2}}) = 23.0:24.5$; 
(5) turbulence velocity $v_{\rm turb}=500:1000$~km~s$^{-1}$; 
(6) chemical abundances as in Table 7.1 of Cloudy documentation.

We first fit the high quality soft state spectra with a model consisting of simple 
blackbody emission absorbed by both neutral material and by the soft state 
photo-ionised plasma model\footnote{We fitted the photo-ionised plasma 
component with the IA model fixing the turbulent velocity either to 
$v_{\rm turb}=500$~km~s$^{-1}$ or $v_{\rm turb}=1000$~km~s$^{-1}$. 
Consistent results are obtained. However, assuming 
$v_{\rm turb}=500$~km~s$^{-1}$ leads to a slightly better fit. For this reason 
we, hereinafter, assume such turbulent velocity for the ionised plasma. }
 (model: \texttt{phabs*IA$_{\rm soft}$*bbody}).
To highlight the effect of the ionised absorber, panel (a) of Fig. \ref{IAdel} 
shows the data and best fit obtained with such a model, once the ionised 
absorber component has been removed. In the same way, panel (b) of 
Fig. \ref{IAdel} shows the data-to-model ratio. 
We show for comparison (panel c) a model consisting of simple blackbody 
emission absorbed by both neutral material and by the \texttt{zxipcf} component 
(model: \texttt{phabs*zxipcf*bbody}).

\begin{figure}
\includegraphics[width=0.37\textwidth,angle=-90]{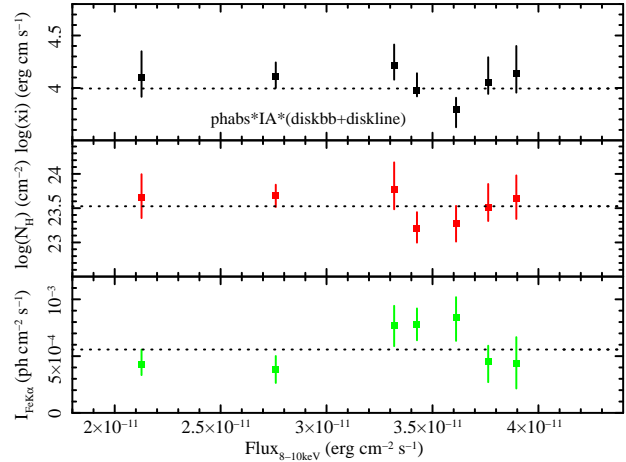}
\caption{Best fit parameter values obtained when fitting the seven high quality soft state spectra with the 
model: \texttt{phabs*IA*(diskbb+diskline)}, as a function of 8--10~keV flux. 
{\it (Top panel:)} Logarithm of the ionisation state of 
the ionised absorber. {\it (Middle panel:)} Logarithm of the column density of the ionised absorber. 
{\it (Lower panel:)} Intensity of the broad Fe~K$\alpha$ emission line. 
}
\label{diskline}
\end{figure}

Adopting the self-consistent absober model from Cloudy (panel d) significantly 
improves the fit ($\chi^2=2458.7$ for $2310$ dof) compared to using the 
\texttt{zxipcf} absorber (compared to $\chi^2=2543.3$ for $2310$ dof). 
In particular the fit is improved at the energy ranges corresponding to the 
Fe~K$\alpha$ and K$\beta$ transitions; strong residuals are no longer present at 
these energies. However, the fit is still statistically unacceptable (null hypothesis probability 
$p$-value=0.016) because significant residuals are still present in the 
$6-7.5$~keV range, as well as an excess of emission above $\sim9$~keV. 
In Section \ref{cont} we observed a scaling of the thermal blackbody temperature 
with luminosity, suggesting an accretion disc origin for this emission, we therefore also attempt to 
model the continuum by replacing the simple blackbody component with a multi-temperature 
disc blackbody component. We obtain a significantly worse fit (see panel d of 
Fig. \ref{IAdel}) with this model (\texttt{phabs*IA*diskbb}; $\chi^2=2589.7$ 
for 2310 dof), and observe strong positive residuals in the range $\sim5.5-8$~keV 
(panel e of Fig. \ref{IAdel}). Note that the disc blackbody model replaces the previous 
excess above $\sim9$~keV with a slight deficit.
To check if the $\sim5.5-8$~keV excess might be produced by a broad 
Fe~K$\alpha$ line, we add a \texttt{diskline} (Fabian et al. 1989) component to the model, 
representing the emission from a line reflected by an accretion disc (panel f of 
Fig. \ref{IAdel}). 
We assume the line energy to be $E_{line}=6.4$~keV (the expected energy from neutral 
iron), the inner and outer disc radii to be $r_{in}=6$~r$_g$, 
$r_{out}=500$~r$_g$ and a disc inclination of $i=80^{\circ}$ 
(the disc inner radius is assumed to be $r_{in}=6$~r$_g\sim12$~km 
as measured in Section 4.2). 
We assume the same disc power-law emissivity index (controlling the radial 
dependence of the emissivity) 
for all soft state spectra and obtain a best-fitting value of $\beta=-2.4$. 
The addition of the \texttt{diskline} component significantly improves
the fit ($\chi^2=2309.6$ for 2301 dof; see Fig. \ref{IAdel}). 
This model now provides a completely statistically acceptable 
description of the data ($p$-value=0.45). 
The broad Fe~K$\alpha$ line component is significantly detected 
(see Fig. \ref{diskline}) within each \xmm\ observation, with an 
equivalent width in the range $EW\sim80-200$~eV. 

We note that the best fitting parameters of the ionised absorption 
component ($N_H$, $\xi$) do not vary significantly if the continuum is 
modelled with either a blackbody or a disc blackbody component, 
and are also independent of whether the broad iron line is added to or 
excluded from the fit. 
The top and middle panels of Fig. \ref{diskline} show the best 
fit column density and ionisation parameter of the ionised absorber model. 

Fig. \ref{diskline} shows that, as previously observed when fitting the absorption 
features with simple Gaussians or the \texttt{zxipcf} model, 
the ionisation level and column density parameters of the Cloudy ionised absorber model 
are consistent with being constant during all the soft state observations. 
No clear trends with luminosity are observed. 

\subsection{Does the wind disappear in the hard state because of ionisation effects?} 

As shown in Section~\ref{SED}, the SED of \axj\ changes dramatically between the soft- and 
the hard state. It is expected that, even if exactly the same absorbing material is present in both 
states, the variation of the source SED will change the ionised plasma ionisation 
state. For example, an increase of the source luminosity in the 8--10~keV band 
is expected to increase (possibly over-ionising) the plasma ionisation, 
changing the intensities of the \Fevc\ and \Fevs\ absorption lines, 
and vice versa for a decrease in the source luminosity.

\begin{figure}
\includegraphics[width=0.37\textwidth,angle=-90]{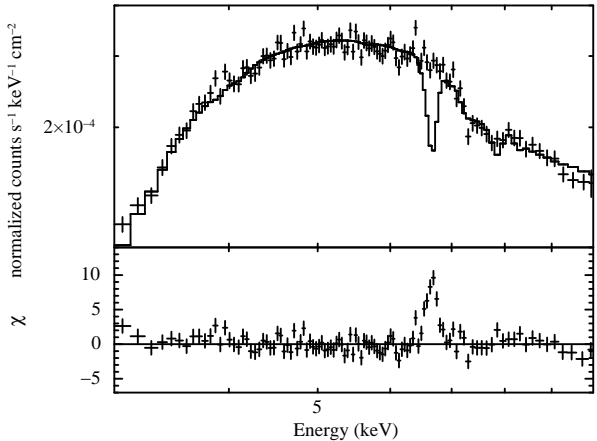}
\caption{Spectrum of the hard state observation 0690441801, compared 
with a model assuming the same ionized plasma observed during the soft state 
(thus, constant $n \times R_0^2$), allowing only the ionisation state to respond 
to changes in the shape and normalisation of the irradiating SED. 
This model and the data are shown in the top panel, and the residuals to the fit 
are shown in the bottom panel. The model is clearly excluded. } 
\label{delHard}
\end{figure}

To check if this is indeed the case, we fist compute for each soft state spectrum, 
from the best fit column density and ionisation parameter, the product $n\times R_0^2$ 
(where $n$ is the number density of the ionised absorbing plasma and $R_0$ is its 
distance from the primary source). In fact, a variation of the SED will change the 
ionisation state of the plasma, but it will leave the product $n\times R_0^2$ constant 
(unless the plasma undergoes a true physical variation). 
Within the soft state observations, we observe this product to remain constant, with 
an average value of $n\times R_0^2=4.8\times10^{32}$~cm$^{-1}$. 
To test if the absorber physically varied during the hard state, we first assume the 
same observed product also for the hard state observations. 
We then derive, given the observed SED, the absorbing plasma ionisation parameter 
for each hard state observation. In particular, the highest statistics hard state 
observation ({\sc obsid}: 0690441801) is expected to have an ionisation parameter 
${\rm log}(\xi/1{\rm ~erg~cm~s^{-1}})=3.19$. We then fit this spectrum with the 
self-consistently ionised absorption model (\texttt{phabs*IA$_{\rm hard}$*diskbb}), 
imposing the expected ionisation parameter and leaving the column density free to 
vary between the lowest and highest value observed during the soft state observation. 
Even for the lowest column density observed during the soft state, the hard state 
ionisation model predicts, at such ionisation state, a very strong \Fevc\ K$\alpha$ 
line (see Fig. \ref{delHard}). The presence of such a line is excluded by the data 
(note that the inclusion of the broad disc line component does not change 
this conclusion). This suggests that ionisation effects are not able to 
explain the observed behaviour and that a true physical variation of the ionised 
absorber between the different states is required. 

\section{Conclusions}

The \xmm\ and \swift\ monitoring observations of the central $\sim15$ arcmin of the Milky Way 
have detected several outbursts from the accreting neutron star system 
\axj, allowing us to measure the spectral evolution during both the soft and hard states.  
Nine \xmm\ observations caught \axj\ in the soft state and three in the hard state. 
Our main conclusions/findings can be summarised as follows:

\begin{itemize} 

\item{}
As is commonly observed in NS-LMXB, the persistent emission of \axj\ during the soft state 
is dominated by a thermal optically thick component, most probably due to multi-temperature 
blackbody emission from the accretion disc, plus blackbody emission from the neutron star 
surface. 
This emission component is observed to vary in temperature with luminosity and to keep its emitting 
area roughly constant. The hard state emission (outside of dipping intervals) is well described 
by a power-law component. The low luminosity of the thermal component is consistent with the 
high-inclination of the system. 

\item{}
The persistent emission in both the soft and hard state is heavily absorbed by a large column 
density of neutral material $N_H\simeq1.9\times10^{23}$~cm$^{-2}$. 
Such a large column density is within a factor 2 of that observed 
towards other Galactic center sources along nearby lines of sight, such as \sgras\ 
and SGR~J1745-2900, and is consistent with remaining constant between all 
\xmm\ observations. This suggests that \axj\ is at (or behind) the Galactic center and that 
most of the obscuring column density is due to the interstellar medium (outside of dipping 
intervals). 

\item{}
Highly significant absorption features due to the \Fevc\ and \Fevs\ K$\alpha$ and K$\beta$ 
lines are detected in the spectra from all nine soft state observations. 
The \Fevc\ and \Fevs\ K$\alpha$ lines 
have typical equivalent widths of $EW\sim-20-35$~eV, very similar to the equivalent widths of the 
corresponding K$\beta$ lines ($EW\sim-15-30$~eV). No absorption lines are observed 
in the hard state (very stringent upper limits are measured; $EW\gsimeq-5-10$~eV). 
This wind-Fe~K absorber vs. state connection is similar to what has been observed in 
\exo, the only accreting neutron star for which the wind-accretion state connection 
has been investigated so far (Ponti et al. 2014). 
Moreover, such behaviour closely resembles what is 
seen in accreting black hole systems, where winds (traced by the same Fe~K absorption 
features) are observed only during the accretion-disc-dominated soft states, and  
disappear during the hard states characterised by jet emission 
(Miller et al. 2008; 2012; Neilsen \& Lee 2009; Ponti et al. 2012). 

\item{} 
We observe the column density of the ionised material 
($N_H\sim2\times10^{23}$~cm$^{-2}$) to be consistent with being constant 
within the seven high-statistics soft state \xmm\ observations. Moreover, we do not observe 
any trends in the \Fevs/\Fevc\ K$\alpha$ and K$\beta$ ratios, 
with the source $8-10$~keV luminosity, between the different soft state observations. 
This might be due to the relatively small range of luminosities spanned by the source 
(less than a factor of 3) and/or be induced by saturation of the lines. 

\item{}
Once the continuum and absorption components are fitted, a broad positive 
residual remains between $\sim5.5-8$~keV. This excess can be reproduced 
by a standard ($EW\sim80-200$~eV) Fe~K$\alpha$ emission line from a 
standard (r$_{in}=6$~r$_g$, r$_{out}=500$~r$_g$) accretion disc seen at 
high-inclination. We note that because the system is highly inclined, the Fe~K$\alpha$ 
line is highly smeared. High signal to noise observations are mandatory 
to reveal the presence of similarly broad emission lines in other high-inclination 
systems.  

\end{itemize}

\section*{Acknowledgments}

The authors wish to thank Jan-Uwe Ness, Ignacio de la Calle and the rest of 
the \xmm\ scheduling team for the enormous support that made the new \xmm\
observations possible. 
GP and TMD acknowledge support via an EU Marie Curie Intra-European fellowship 
under contract no. FP-PEOPLE-2012-IEF-331095 and FP-PEOPLE-2011-IEF-301355, 
respectively. The GC \xmm\ monitoring project is partially supported by the 
Bundesministerium f\"{u}r Wirtschaft und Technologie/Deutsches Zentrum f\"{u}r Luft- 
und Raumfahrt (BMWI/DLR, FKZ 50 OR 1408) and the Max Planck Society. 
This project was funded in part by European Research Council Advanced 
Grant 267697 "4 $\pi$ sky: Extreme Astrophysics with Revolutionary Radio Telescopes". 
DH acknowledges support from Chandra X-ray Observatory (CXO) Award Number 
GO3-14121X, operated by the Smithsonian Astrophysical Observatory for and on behalf 
of NASA under contract NAS8-03060, and also by NASA Swift grant NNX14AC30G.
CH is supported by an NSERC Discovery Grant, and an Ingenuity New Faculty Award.
ND is supported by NASA through Hubble Postdoctoral Fellowship grant number 
HST-HF-51287.01-A from the Space Telescope Science Institute. 
The scientific results reported in this article are based on observations obtained 
with \xmm, \swift\ and \nustar.

\begin{table*}
\begin{center}
\small
\begin{tabular}{ c r c r r c r r c c c c c c c }
\hline
\hline
\multicolumn{2}{c}{\textbf{\emph{XMM-Newton}} } \\
 {\sc OBSID}  & Rev   &         START (UTC) & EXP  &CL EXP &STATE& F$_{3-6}$& F$_{6-10}$& F$_{8-10}$&Threshold\\
                      &          &                                & (ks)   & (ks)      &            &\multicolumn{3}{c}{($10^{-12}$~erg~cm$^{-2}$~s$^{-1}$)} & (ct~s$^{-1}$)   \\
\hline
  0690441801 & 2622  & 2014-04-03 05:06:00 &  86.5 & 69.1 & H &5.95&10.1&4.75&3.5/0.24/1.5/0.65/0.5/1.5\\
  0724210501 & 2525  & 2013-09-22 21:15:49 &  43.9 & 30.0 & S &55.8&59.8&21.7&10/1.1/1.4/5/3.8/1.5 \\
  0700980101 & 2519  & 2013-09-10 03:30:45 &  38.7 & 30.6 & S &78.3&91.5&33.5&13/1.5/1.4/7/5/1.5   \\
  0724210201 & 2514  & 2013-08-30 20:13:12 &  58.5 & 45.7 & S &72.9&91.5&34.5&13/1.6/1.5/7/4.5/1.4 \\
  0694641101 & 2343  & 2012-09-24 10:16:44 &  41.9 & 35.6 & Q &$\flat$&$\flat$&     & 0.8                               \\
  0694641001 & 2343  & 2012-09-23 20:20:07 &  47.9 & 40.8 & Q &$\flat$&$\flat$&     & 0.85                               \\
  0694640301 & 2331  & 2012-08-31 11:20:26 &  41.9 & 35.5 & Q &$\flat$&$\flat$&     &  0.75                              \\
  0674601001 & 2249  & 2012-03-21 03:30:40 &23.9$\P$&17.2&Q&$\flat$&$\flat$&     &  0.7                              \\
  0674600801 & 2248  & 2012-03-19 03:52:38 &22.9$\P$&15.0&Q&$\flat$&$\flat$&     &  0.7                              \\
  0674601101 & 2247  & 2012-03-17 02:30:16 &28.0$\P$& 9.8 &Q&$\flat$&$\flat$&     &  0.6                              \\
  0674600701 & 2246  & 2012-03-15 04:47:06 &  15.9 &0.1\ddag&Q&      &         &       &  0.8                              \\
  0674600601 & 2245  & 2012-03-13 03:52:36 &21.5$\P$& 7.4 &Q&$\flat$&$\flat$&     &  0.9                              \\
  0658600201 & 2148  & 2011-09-01 20:03:48 &53.2$\P$&36.5&Q&$\flat$&$\flat$&     &  0.9                              \\
  0658600101 & 2148  & 2011-08-31 23:14:23 &49.9$\P$&41.9&Q&$\flat$&$\flat$&     &  0.9                              \\
  0604301001 & 2073  & 2011-04-05 07:09:33 &50.7$\P$&28.7&Q&$\flat$&$\flat$&     &  0.6                              \\
  0604300901 & 2072  & 2011-04-03 07:52:07 &46.9$\P$&16.8&Q&$\flat$&$\flat$&     &  0.6                              \\
  0604300801 & 2071  & 2011-04-01 07:48:13 &48.8$\P$&29.5&Q&$\flat$&$\flat$&     &  0.7                              \\
  0604300701 & 2070  & 2011-03-30 07:44:39 &48.9$\P$&28.5&Q&$\flat$&$\flat$&     &  0.6                              \\
  0604300601 & 2069  & 2011-03-28 07:49:58 &48.8$\P$&24.9&Q&$\flat$&$\flat$&     &  0.6                              \\
  0554750601 & 1707  & 2009-04-05 02:17:13 &39.1$\P$&27.3&Q&$\flat$&$\flat$&     &  0.6                              \\
  0554750501 & 1706  & 2009-04-03 01:33:27 &44.3$\P$&33.4&Q&$\flat$&$\flat$&     &  0.65                              \\
  0554750401 & 1705  & 2009-04-01 00:55:25 &39.9$\P$&27.5&Q&$\flat$&$\flat$&     &  0.65                              \\
  0511000401 & 1610  & 2008-09-23 15:15:50 &    6.9 &   4.4 & D &0.15&0.23&0.10&                             \\
  0505670101 & 1518  & 2008-03-23 14:59:43 &105.7 & 65.2 & S &64.6&75.7&27.9&11/1.3/1.3/5/4/1.6 \\
  0511000301 & 1508  & 2008-03-03 23:25:56 &    6.9 &   4.4 & S &72.4&85.0&31.1&12/1.5/1.4/5/3.7/20\\
  0504940201 & 1418  & 2007-09-06 10:05:50 &  13.0 &   8.5 & S &54.7&60.5&22.2&12/1.2/1.5/4.2/3.2/2.5\\
  0402430401 & 1340  & 2007-04-03 14:32:24 & 105.7 & 38.6 & S &74.9&95.6&36.4&12/1.5/1.4/6/4.3/1.5  \\
  0402430301 & 1339  & 2007-04-01 14:45:02 & 105.4 & 59.6 & S &78.2&99.0&37.9&12/1.4/1.3/6/5/2      \\
  0402430701 & 1338  & 2007-03-30 21:05:17 &  34.2 & 24.4 & S  &84.6&105&39.8&12/1.5/1.4/6/4.3/2    \\
  0302884001 & 1236  & 2006-09-08 16:56:48 &    6.9 &   3.4 & D  &0.37&0.68&0.33& 2/0/1.1/0.08/0.07/0.6 \\
  0302882601 & 1139  & 2006-02-27 04:04:34 &    6.9 &   4.0 & H  &8.02&13.7&6.28& 4/0.001/1.2/0.9/0.6/5 \\
  0202670801 &  867  & 2004-09-02 03:01:39 & 135.2 & 75.2 & D  &0.21&0.38&0.18& 0.65                           \\
  0202670701 &  866  & 2004-08-31 03:12:01 & 135.2 & 81.8 & D  &0.20&0.28&0.12& 0.6                           \\
  0202670601 &  789  & 2004-03-30 14:46:36 & 134.4 & 27.2 & D  &0.37&0.68&0.33& 0.6                           \\
  0202670501 &  788  & 2004-03-28 15:03:52 & 133.0 & 13.7 & D  &0.23&0.44&0.22& 0.6                            \\
  0111350301 &  516  & 2002-10-03 06:54:11 &  17.3 &    7.6 & D  &0.44&0.87&0.43& 0.55                           \\
  0111350101 &  406  & 2002-02-26 03:16:43 &  52.8 &  35.8 & D  &0.11&0.19&0.09& 0.6                           \\
  0112972101 &  318  & 2001-09-04 01:20:42 &  26.7 &  18.2 & Q  &$\flat$&$\flat$&   & 0.55                           \\
\hline
\end{tabular}
\caption{
A list of all \xmm\ observations considered in this work. 
The columns of the table report the \xmm\ {\sc OBSID}, the \xmm\ revolution, 
the observation start date and time, the observation duration and the {\sc EPIC-pn} 
exposure time after cleaning, the source state (H=hard state; S=soft state; 
D=detected (the source is detected but is still in quiescence); Q=quiescent). 
The following columns give the $3-6$, $6-10$ and $8-10$~keV 
observed (absorbed) fluxes in units of $10^{-12}$~erg~cm$^{-2}$~s$^{-1}$. 
$\flat$ The fluxes of \axj\ during the observations in 
quiescence are $F_{\rm 3-6~keV} < 5\times10^{-14}$ and 
$F_{\rm 6-10~keV} < 8\times10^{-14}$~\ergcms.
The last column shows, in order, the count-rate thresholds applied to select bursting, 
eclipsing and intense dipping periods, the hard and soft count rates and the 
threshold to select out intense particle activity periods. 
A more exhaustive description of the data reduction and cleaning is provided in Section 2.1. 
$\P$ Datasets used to compute the source background for the outburst observations. 
\ddag Exposure too short to compute a meaningful upper limit on the flux of \axj\ 
(no source emission is observed even in the uncleaned data). }
\label{data}
\end{center}
\end{table*} 

\begin{table*}
\begin{center}
\small
\begin{tabular}{ c c c c | c c c | c c c c c c c}
\hline
\hline
\multicolumn{2}{c}{\textbf{\emph{XMM-Newton}}}&\multicolumn{3}{c}{\textbf{\emph{phabs*power-law}}}&\multicolumn{4}{c}{\textbf{\emph{phabs*diskbb}}}&$\Delta\chi^2$~~ \\
  OBSID   &STATE& N$_{H}$          & $\Gamma$       & $\chi^2$/dof & N$_{H}$          & T$_{DBB}$     & $\chi^2$/dof  & $\chi^2$/dof  & PL-BB\\
                &   & ($10^{22}$~cm$^{-2}$) &                    &              &($10^{22}$~cm$^{-2}$)& (keV)        &                       & (no Fe~K) \\         
\hline
  0690441801 & H & $18.9\pm0.5$ & $1.85\pm0.05$ & 1347.5/1259 & $15.6\pm0.4$   & $3.3\pm0.1$ & 1372.9/1034 &                  & -63.3  \\
  0724210501 & S & $24.1\pm0.4$ & $3.07\pm0.04$ &   639.8/332   & $17.3\pm0.3$ & $1.95\pm0.03$ & 425.4/332   & 202.7/192 & 223.1 \\
  0700980101 & S & $25.8\pm0.4$ & $2.99\pm0.03$ &   908.9/332   & $19.1\pm0.3$ & $2.00\pm0.02$ & 500.6/332   & 200.9/192 & 469.5 \\
  0724210201 & S & $25.6\pm0.3$ & $2.83\pm0.03$ &   986.7/332   & $19.2\pm0.2$ & $2.12\pm0.02$ & 506.4/332   & 208.8/192 & 528.4 \\
  0505670101 & S & $25.8\pm0.3$ & $2.97\pm0.03$ &   952.8/332   & $19.2\pm0.2$ & $2.00\pm0.02$ & 570.2/332   & 198.1/192 & 388.8 \\
  0511000301 & S & $25.7\pm1.0$ & $2.96\pm0.10$ &   374.9/332   & $19.2\pm0.7$ & $2.01\pm0.07$ & 346.6/332   & 208.0/192 &  28.5 \\
  0504940201 & S & $25.5\pm0.7$ & $3.09\pm0.08$ &   429.0/332   & $18.6\pm0.6$ & $1.94\pm0.05$ & 374.1/332   & 194.1/192 &  55.8 \\
  0402430401 & S & $26.3\pm0.3$ & $2.82\pm0.03$ &   813.0/332   & $20.0\pm0.2$ & $2.11\pm0.02$ & 499.4/332   & 193.0/193 & 349.6 \\
  0402430301 & S & $24.9\pm0.3$ & $2.76\pm0.03$ &   965.1/332   & $18.9\pm0.2$ & $2.16\pm0.02$ & 577.1/332   & 202.6/192 & 349.5 \\
  0402430701 & S & $25.8\pm0.4$ & $2.86\pm0.04$ &   592.0/332   & $19.4\pm0.3$ & $2.08\pm0.03$ & 405.9/332   & 163.8/192 & 192.2 \\
  0302882601 & H & $21.8\pm2.1$ & $1.99\pm0.20$ &   247.1/246   & $17.8\pm1.6$ & $3.04\pm0.38$ & 254.3/246   &                  &   -2.5 \\
\hline
\end{tabular}
\caption{Best-fitting parameters for the \xmm\ outburst observations when  
fitted with simple phenomenological continuum models. 
The first two columns report the \xmm\ {\sc OBSID} and the source state (see Section \ref{State}).
The following three columns report the best fitting neutral column density and spectral index
parameters and the $\chi^2/$dof obtained for an absorbed power-law model (\texttt{phabs*power}).
The following three columns report the neutral column density, the disc blackbody temperature, 
the minimum $\chi^2$ and the dof obtained for an absorbed disc blackbody model (\texttt{phabs*diskbb}). 
The ninth column shows the $\chi^2$ and the dof obtained for an absorbed disc blackbody 
model (\texttt{phabs*diskbb}), once the Fe~K band ($5.5<E<8.5$~keV) is excluded from the fit. 
The last column reports the difference in $\chi^2$ between the best-fitting absorbed power-law model 
minus the best-fitting absorbed disc blackbody model. 
} 
\label{fit}
\end{center}
\end{table*} 

\begin{table*}
\begin{center}
\scriptsize
\begin{tabular}{ c c c c c c c c c c c }
\hline
\hline
\multicolumn{2}{c}{\textbf{\emph{XMM-Newton}} } & \multicolumn{8}{c}{\textbf{\texttt{phabs*bbody*(gaus+gaus+gaus+gaus) } }} \\
  OBSID &       & \multicolumn{2}{c}{\Fevc~K$\alpha$} & \multicolumn{2}{c}{\Fevs~K$\alpha$} & \multicolumn{2}{c}{\Fevc~K$\beta$} & \multicolumn{2}{c}{\Fevs~K$\beta$} \\
              &           & E (keV) &EW (eV) &E (keV) &EW (eV) &E (keV) &EW (eV) &E (keV) & EW (eV) & $\chi^2/dof$ \\
\hline
  0690441801  & H\dag &                          & $>-7$       &                          &  $>-13$          \\
  0724210501 & S & $6.69\pm0.05$ & $-17.6\pm7.5$ & $6.94\pm0.02$ & $-33.3\pm8.1$ & $7.88\pm0.05$ & $-37.9\pm10.8$ & $8.17\pm0.02$ & $-16.2\pm13.0$ & 319.1/324 \\
  0700980101 & S & $6.73\pm0.04$ & $-16.9\pm6.0$ & $6.96\pm0.03$ & $-31.0\pm6.7$ & $7.93\pm0.04$ & $-16.5\pm9.0$   & $8.22\pm0.02$ & $-23.2\pm9.8$   & 327.2/324 \\
  0724210201 & S & $6.75\pm0.04$ & $-13.5\pm4.9$ & $6.98\pm0.03$ & $-20.1\pm5.5$ & $7.88\pm0.04$ & $-17.2\pm7.5$   & $8.24\pm0.03$ & $-17.5\pm8.2$   & 379.8/324 \\
  0505670101 & S & $6.64\pm0.02$ & $-28.7\pm4.9$ & $6.95\pm0.02$ & $-32.4\pm5.6$ & $7.83\pm0.02$ & $-25.2\pm7.8$   & $8.17\pm0.02$ & $-34.0\pm8.7$   & 348.3/324 \\
  0402430401 & S & $6.68\pm0.02$ & $-33.3\pm5.8$ & $6.99\pm0.03$ & $-23.2\pm6.9$ & $7.87\pm0.02$ & $-26.3\pm9.2$   & $8.17\pm0.03$ & $-24.8\pm10.3$ & 321.8/324 \\
  0402430301 & S & $6.65\pm0.02$ & $-26.9\pm4.8$ & $6.94\pm0.02$ & $-35.9\pm5.2$ & $7.78\pm0.02$ & $-11.6\pm7.1$   & $8.17\pm0.02$ & $-25.2\pm8.1$   & 339.3/324 \\
  0402430701 & S & $6.68\pm0.04$ & $-22.8\pm7.1$ & $6.99\pm0.02$ & $-39.1\pm8.0$ & $7.80\pm0.04$ & $-32.2\pm10.1$ & $8.19\pm0.02$ & $-27.3\pm12.7$ & 287.4/324 \\
\hline
\end{tabular}
\caption{Best-fitting model parameters obtained for the high statistics \xmm\ soft state spectra 
when fitting with the absorbed blackbody plus four Gaussian line model 
{\texttt{phabs*bbody*(gauss+gauss+gauss+gauss)}}. 
The first two columns report the  \xmm\ {\sc OBSID} and the state of the source. 
The following 8 columns show the best-fitting energy and equivalent width 
(or upper limit) for the \Fevc~K$\alpha$, \Fevs~K$\alpha$, \Fevc~K$\beta$ and 
\Fevs~K$\beta$ lines, respectively. All lines are significantly detected in each 
high quality soft state spectrum. Only upper limits on the equivalent widths are observed during 
the hard state observations (only the longer hard state \xmm\ observation is reported here). 
The last column reports the best-fitting $\chi^2/$dof for the absorbed blackbody 
model plus four Gaussians. 
\dag The hard state spectrum is fitted with an absorbed power-law model. 
} 
\label{fitlines}
\end{center}
\end{table*} 

\end{document}